\documentclass[a4paper,11pt]{article}
\pdfoutput=1 % if your are submitting a pdflatex (i.e. if you have
\usepackage{jcappub} % for details on the use of the package, please
\usepackage[T1]{fontenc} % if needed

\usepackage{float}
\usepackage[caption = true]{subfig}             
\usepackage{diffcoeff} 
\usepackage{amsmath,bm}
\usepackage[italicdiff]{physics}
\usepackage{hyperref}
\usepackage{multicol}

\title{\boldmath Effect of Magnetised Discontinuity on Diffusive Shock Acceleration}

\author[a]{Anshuman Verma}
\author[a]{Saksham Chandna}
\author[a]{Divyansh Tripathi}
\author[a]{Ritam Mallick}
\affiliation[a]{Indian Institute of Science Education and Research Bhopal, Bhauri, Bhopal, \\ Madhya Pradesh, India}
\emailAdd{anshuman18@iiserb.ac.in}
\emailAdd{sakshamc537@gmail.com}
\emailAdd{divyansh20@iiserb.ac.in}
\emailAdd{mallick@iiserb.ac.in}
\abstract{We investigate the impact of magnetic fields and diffusion mechanisms on the energy spectra of particles accelerated via diffusive shock acceleration. We analyse magnetised shock jump conditions and demonstrate how magnetisation and angular dependence modify upstream and downstream velocities, which enter the transport equation within a Monte Carlo simulation framework. We consider constant, momentum-dependent, and pitch-angle-dependent diffusion coefficients to assess their influence on particle acceleration. Our results show that magnetic fields enhance particle confinement and facilitate more efficient energy gain. In the absence of magnetisation, particle spectra tend to be steeper due to rapid escape and weaker scattering effects, whereas magnetised shocks systematically produce flatter spectra across all diffusion models. Among them, pitch-angle-dependent diffusion leads to the strongest spectral flattening, underscoring its role in sustaining extended acceleration. It is also seen that an increased upstream pressure, associated with enhanced magnetic turbulence, broadens the spectral range by improving particle scattering efficiency and enabling multiple shock crossings. As the shock inclination angle increases, the velocity contrast between upstream and downstream regions diminishes, modulating the spatial extent of the acceleration zone. Notably, pitch-angle-dependent diffusion remains robust under varying shock conditions, ensuring sustained acceleration.}

\begin{document}
\maketitle
\flushbottom

\section{Introduction}\label{sec:1}
Shock waves are a common occurrence in astrophysical scenarios. From solar bow shocks to extragalactic shocks in supernovae, they are responsible for many astrophysical phenomena. In astrophysics, relativistic treatment of shocks is more appropriate \citep{taub, anile, ansh1}. In some cases, hydrodynamic shocks are also associated with the magnetic field, which is relevant for studying environments having significant magnetic contributions \citep{mallick_2014a, Mishustin_2014,rm_singh19,singh_2019, Mallick19, Mallick_2020}. However, the magnetic field complicates the study, and one has to shift to a special frame, the de Hoffmann frame, to simplify the problem and make the analysis tractable \citep{de-hoffmann}. Assuming the fluid has infinite conductivity, and the flow velocity is parallel to the magnetic field in the frame, preventing self-induction of the magnetic field for the fluid at rest. The assumption is applicable to astrophysical shocks, where the discontinuity interface can be considered infinitesimal compared to the extent of the shock.

In this analysis, we primarily focus on studying the effect of magnetohydrodynamic shocks on particle acceleration. The theory of particle acceleration through shock waves emerged in the late $ 1970$ \citep{Aflven43, Fermi49, Krymskii77} and has since become the dominant framework for understanding astrophysical particle acceleration. A universal feature of high-energy astrophysical phenomena (such as Supernova remnants (SNRs), Gamma-Ray bursts (GRBs) and Active Galactic Nuclei (AGNs)) is the release of highly-relativistic particles from them \citep{Bell78, BlandfordOstriker78, Kuzur_2020}. These particles are accelerated to high energies near these energetic environments and emitted, forming cosmic radiation \citep{Prishchep81, Berezhko02}. These particles are detected directly as cosmic radiation or indirectly as non-thermal radiation. The synchrotron emission and inverse Compton radiation from the accelerated electrons exhibit a power-law energy spectrum in specific energy ranges \citep{Blandford87}. The cosmic ray spectrum obtained through various experimental ventures is displayed in many literature \citep{Longair11}. The ``knee" of the spectrum is observed around the $PeV$ energy range, where a steepening of the spectral curve is seen. This feature probably corresponds to the upper end of Galactic Cosmic Rays (GCRs) \citep{Gaisser2006, Liu-2024}. In lower energy ranges, from $GeV$ to $TeV$, most of the observed GCRs seem to originate in regions around a particular class of supernova explosions, which involve rapid evolution and collapse of massive progenitor stars (Wolf-Rayet stars) \citep{Biermann19}. The highest energies for GCRs are thought to emerge from strong magnetic fields in the acceleration region near Sedov-Taylor blast waves in SNRs \citep{Bell12}. Beyond the "knee" of the spectrum, extra-galactic sources are thought to be primarily responsible. Our focus is mainly on the power-law energy spectrum observed before the "knee" of the spectrum, where sources and production mechanisms are more easily identifiable using the diffusive shock acceleration (DSA) model \citep{Aartsen04,apeli4}. For $E<10^{15}eV$, the power-law energy distribution: $N_{obs}(E) dE \propto E^{-2.7} dE$ is observed \citep{Adriani14,Aguilar15,Neronov17}. In non-thermal radio sources like SNRs and radio galaxies, synchrotron emission by $GeV$ electrons also gives a similar energy spectrum, where the power law exponent $s$ ranges between $s \approx 2.2-3$ \citep{Yang14,acero15,neronov15}. The source spectrum is thought to arise from the competition between the energy gained per shock-crossing cycle (in Sedov-Taylor blasts) and the contribution from the probability of plasma particles escaping the acceleration region after each collision cycle \citep{Achterberg01, Gaisser2006}.

The DSA mechanism, initially proposed by Fermi, involves stochastic scattering of plasma particles by Alfvén waves near collisionless shock fronts. This scattering confines particles near the shock, allowing for prolonged acceleration \citep{Fermi49,Bell78,Drury83,kirkdendy01}. DSA occurs when the relative upstream fluid velocity exceeds the downstream velocity (in the rest frame of the shock). As particles cross the shock, they encounter the relative motion of fluid (across the shock front), leading to energy gains through Fermi acceleration \citep{Bell04}. The interplay between energy gain from shock crossings and the likelihood of particles moving downstream generates a power-law spectrum. Magnetic field irregularities around the shock confine particles near the shock front. In non-relativistic cases, an isotropic particle velocity distribution is assumed due to scattering, forming the basis for many DSA analyses \citep{Bell04,Bell12}. In relativistic shocks, anisotropic particle velocity distributions arise due to comparable velocities of the bulk plasma and individual particle velocities \citep{apjac7049bib2,apjac7049bib64,apjac7049bib65,apjac7049bib1,apjac7049bib66}. Additionally, the orientation of magnetic field lines relative to the shock normal influences the spectral index. Thus, for relativistic shock acceleration, semi-analytic or numerical models are suitable for determining the spectral index \citep{Kirk1987I, Kirk1987II, Axford77, Kirk1987I, Kirk1987II, Ballard91, Achterberg01, KirkAchterberg00, Gallant02}.

The DSA problem can be tackled through two main approaches: analytical and numerical. The analytical approach, using a shockwave, analytically derives a spectral index value of $s = 2$ for non-relativistic strong parallel shocks in a diffuse medium \citep{Bell78, Alan89}. The numerical approach solves the transport equation for upstream and downstream regions of non-relativistic shocks, matching solutions at the shock discontinuity \citep{BlandfordOstriker78}. 
For relativistic shock scenarios, a numerical approach is considered that takes into account the competition between particle acceleration in converging flows of scattering centres and escape into the downstream region, resulting in power-law spectra \citep{Peacock81, Kirk1987I, Kirk1987II}.
Gallant reviewed theoretical aspects of particle acceleration with ultra-relativistic and moderately relativistic shocks \citep{Gallant02}.
However, these theoretical approaches often struggle to comprehensively describe the spectral index values in more realistic DSA cases. A realistic treatment of the DSA problem necessitates solving the corresponding Fokker-Planck-type equation for particles \citep{fokker89, Arzner04, Arzner06}. 

In this paper, we analyse the effect of magnetic fields on DSA. We begin by formulating the dynamics of relativistic shock waves in a magnetised medium, extending the established frameworkAssuming the shock front is perpendicular to the flow, and the discontinuity's width is negligible compared to the actual system, we have only a single discontinuity across which the flow variables are discontinuous. When utilised in the context of DSA, it serves as a key input for Monte Carlo simulations that model particle transport equations. Through this approach, we obtain the energy spectra of accelerated particles and determine their spectral indices. A distinguishing feature of this methodology is that it inherently accounts for the influence of the magnetic field and the obliquity of the shock waves in DSA by incorporating the derived velocity profiles.
This paper is structured as follows:  In section \ref{sec:2}, we give a detailed description of the formalism of magnetised shock discontinuity, the EoS, the transport equation and the energy spectrum. Next, in Section \ref{sec:5}, we present our results, while Section \ref{sec:6} provides a summary of our key findings and conclusions.

\section{Formalism}\label{sec:2}
\subsection{Shock Waves}
Assuming the shock front is perpendicular to the flow, and the discontinuity's width is negligible compared to the actual system, thus having only a single discontinuity across which the flow variables are discontinuous. Denoting the two sides of the shock discontinuity as "a" and "b," the difference of a thermodynamic quantity (Q) across the front is given by [Q] = $Q_a$ - $Q_b$. The discontinuous surface is denoted by $\Sigma$, having a unit normal vector (NV), $\Lambda^\mu$, in space-time (ST). The shock discontinuity can be either a space-like (SL) or a time-like (TL) shock according to the normalisation condition given \cite{taub_1978,Csernai94,bugaev_2000,zhang_2014,mallick_2014a, Mallick19,ansh1,ansh2} as,
\begin{equation*}
    \Lambda^{\mu}\Lambda_{\mu}  = \begin{cases}
                                        -1, & \text{For SL Hypersurface $\Sigma$} \\
                                        +1, & \text{For TL Hypersurface $\Sigma$}
                                    \end{cases}
\end{equation*}

%\subsubsection{Special Relativistic Shock Waves}
%The equations for the Special Relativity (SR) shock waves are formulated based on the energy-momentum tensor ($T^{\mu\nu}$). 
The line element in flat ST is given by

\begin{equation}
    ds^2= \eta_{\mu\nu}dx^{\mu}dx^{\nu}=-dt^2 + dr^2 + r^{2}d\Omega^{2},
\end{equation}
where $\eta_{\mu\nu} $ is flat ST metric in spherical polar coordinates. 
Considering a perfect fluid, the stress-energy tensor is given by
\begin{equation}
    T^{\mu\nu} = wu^{\mu}u^{\nu} + p\eta^{\mu\nu}
\end{equation}
where, $w$ (the enthalpy) is equal to the $e$ (energy density) + $p$ (pressure), $u^{\mu}=(\gamma,\gamma v,0,0)$ is the four-velocity vector with $\gamma = \frac{1}{\sqrt{1 - v^{2}}}$ and its norm is given by $\eta_{\mu\nu}u^{\mu}u^{\nu} = -1$.

The jump condition across the shock front can be written in general (described well in \cite{taub, anile,ansh1,ansh2}) as,

\begin{align}
    T_{a}^{\mu\nu}\Lambda_{\nu} & = T_{b}^{\mu\nu}\Lambda_{\nu} \label{J1}\\
    n_{a}u_{a}^{\mu}\Lambda_{\mu} &= n_{b}u_{b}^{\mu}\Lambda_{\mu}. \label{J2}
\end{align}

%Using this condition, one can derive the Rankine-Hugoniot jump conditions \cite{}. 

%\subsubsection{Effect of magnetic field}
%%%===================================

The magnetic field affects both the electromagnetic stress-energy tensor and the EoS. 
%First, we discuss the effect of the magnetic field in the jump condition, which comes from the stress-energy tensor.
To simplify the problem, we assume an ideal, infinitely conducting fluid in the conservation equation for Magnetohydrodynamics (MHD) shocks, where the energy-momentum tensor incorporates contributions from both matter and magnetic fields \citep{debaes_1997,prakash_2000,rm_singh19}. Consequently, the electric field is assumed to be negligible. The total energy-momentum tensor is expressed as,

\begin{equation}
    T^{\mu\nu} = T^{\mu\nu}_{M} + T^{\mu\nu}_{B} \label{mag_T}
\end{equation}
where "M" represents the matter part of the tensor, and "B" represents the magnetic part. The magnetic part of the tensor is defined as

\begin{align}
    T^{00}_{B} &= \frac{B^{2}}{8\pi} \\
    T^{ij}_{B} &= \frac{B^{2}}{8\pi}\delta^{ij} - \frac{B^i B^j}{4\pi} 
\end{align}
where $B^i$ is the magnetic field vector.

\begin{figure}[ht]
\hspace{0.5cm}
  \includegraphics[scale=1.0]{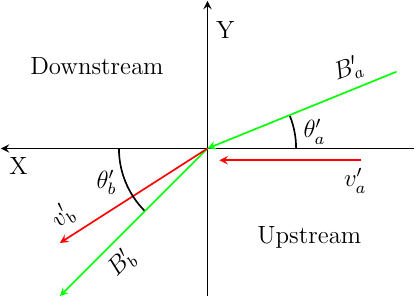} 
  \includegraphics[scale=1.0]{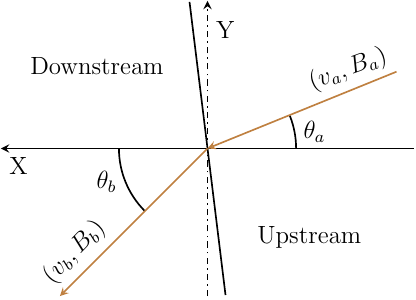} 
  \caption{Schematic diagram illustrating a moving shock discontinuity in both the NI-frame (primed) and the HT-frame (unprimed). In both reference frames, the shock front propagates from left to right. The matter velocities on either side of the shock, denoted as \( v_a \) and \( v_b \) (with primed/unprimed variations), are oriented at angles \( \theta_a \) (incident angle, primed/unprimed) and \( \theta_b \) (reflected angle, primed/unprimed) relative to the shock normal. In the HT-frame, the velocity and magnetic field vectors become collinear, simplifying the analysis of the problem.}\label{f1:frame}
\end{figure}

Having the shock conservation condition, i.e., the relativistic Rankine-Hugoniot condition, can be quite challenging for MHD shocks. One approach to making this problem more manageable is to adopt the De Hoffmann Teller frame (HT) \citep{de-hoffmann}. The HT frame (see figure \ref{f1:frame}) serves as the shock rest frame, where there are no drift electric fields resulting from the cross product of the velocity vector (v) and the magnetic field vector (B). 
 \begin{align}
    \vec{v} \times \vec{B} = 0 \label{HTc}
 \end{align}

Consequently, for subluminal flows, the HT frame is a natural choice, as it significantly simplifies the conservation conditions. We can transfer from the HT frame to the normal incidence (NI) frame (or lab frame) with a simple transformation, as shown in the figure and explained in Appendix \ref{A1}.
%In addition to the system of conservation equations, we also incorporate an EoS to describe the material properties in the upstream and downstream regions (described in the next subsection \ref{EoS}). 
Additionally, we assume that the fluid flows along the magnetic field lines in the HT frame. The four matter jump conditions encompass the conservation of baryon number, momentum (in two components), and energy across the shock front \citep{mallick_2014a,mallick_2014b,rm_singh19}.
\begin{itemize}
    \item For an SL discontinuity, the jump conditions can be written as 
\end{itemize}     
    \begin{align}
      T_{a}^{01} &=  T_{b}^{01} \nonumber\\
%\text {Conservation of Energy flux} \nonumber \\
     \Rightarrow w_{a}\gamma_{a}^{2}v_{ax}  &= w_{b}\gamma_{b}^{2}v_{bx} \\ 
     T_{a}^{11} &=  T_{b}^{11} \nonumber \\
%\text {Conservation of momentum flux} \nonumber \\
      \Rightarrow w_{a}\gamma_{a}^{2}v_{ax}^{2} + p_{a} + \frac{B_{ay}^2}{8\pi} &= w_{b}\gamma_{b}^{2}v_{bx}^{2} + p_{b} + \frac{B_{by}^2}{8\pi}\\ 
       T_{a}^{21} &=  T_{b}^{21} \nonumber \\
%\text {Conservation of momentum flux} \nonumber \\
      \Rightarrow w_{a}\gamma_{a}^{2}v_{ax}v_{ay} - \frac{B_{ax}B_{ay}}{4\pi} &= w_{b}\gamma_{b}^{2}v_{bx}v_{by} - \frac{B_{bx}B_{by}}{4\pi}\\ 
     n_{a}u_{a}^{1}  &= n_{b}u_{b}^{1} \nonumber \\
%\text {Conservation of particle number flux} \nonumber \\
      \Rightarrow n_{a}v_{ax}\gamma_{a} &= n_{b}v_{bx}\gamma_{b} 
    \end{align}  

The equation governing the electromagnetic jump condition is
\begin{align}
    \nabla \cdot \Vec{B} &= 0 \\
    \nabla\cross \Vec{E} &= 0
\end{align}
where the last equation arises trivially because $\Vec{E} = 0$ holds everywhere \citep{ferraro_1954,konno_1999}. We also assume that the x-direction defines the normal to the shock front. The magnetic field is constant and lies in the x-y plane. Therefore, the velocities in the x and y directions are given by $v_{x}$ and $v_{y}$, respectively. Similarly, the magnetic fields in the x and y directions are given by $B_{x}$ and $B_{y}$. The Lorentz factor is defined as
\begin{equation*}
    \gamma_i = (1 - v_{ix}^{2} - v_{iy}^{2})^{-1/2}
\end{equation*}
where $i=a,b$.
By using equation \ref{HTc} for the HT frame, we can also obtain,

\begin{align}
    \frac{v_{ay}}{v_{ax}} = \frac{B_{ay}}{B_{ax}} \equiv tan \theta_a \label{HT1}\\ 
    \frac{v_{by}}{v_{bx}} = \frac{B_{by}}{B_{bx}} \equiv tan \theta_b \label{HT2}
\end{align}
With the assumption of infinite conductivity, the electric field is zero. The Maxwell equation $\nabla. B = 0$ defines that there are no magnetic monopoles, resulting in
\begin{equation}
    B_{ax} = B_{bx} 
\end{equation}

Using the jump conditions, one can further derive the combustion adiabat (CA) equations for magnetised SL shock \citep{mallick_2014a,mallick_2014b,rm_singh19}. 
The CA equation is particularly interesting when the EoS differs on either side of the front. For DSA, the EoS on either side of the shock front remains the same, and the jump conditions are sufficient. Still, for completeness, we show the CA equation, which can be derived from the jump conditions.
The CA equation is a velocity-independent scalar equation, making it easily solvable. 
%It is derived from the general form of the jump conditions presented in Eqs. \ref{J1} and \ref{J2}. 
The energy-momentum tensor conservation condition (defined in Eq. \ref{mag_T}) can be solved to systematically derive the equation (detailed in \citep{mallick_2014a,rm_singh19}),

\begin{itemize}
    \item For Space-like Shocks, the CA equation is expressed as
\begin{align}
   &\Bigg(\frac{w_{a}^2}{n_{a}^{2}} - \frac{w_{b}^2}{n_{b}^{2}}\Bigg)\Bigg(\frac{w_{a}}{n_{a}^{2}} - \frac{w_{b}}{n_{b}^{2}}\Bigg) + \Bigg(p_{b} - p_{a} + \frac{B_{by}^2 - B_{ay}^2}{8\pi}\Bigg) \Bigg(\frac{w_{a}^2}{n_{a}^{4}h_a} - \frac{w_{b}^2}{n_{b}^{4}h_b}\Bigg) = 0. 
\end{align}
where $h_a = Cos^2_{\theta_a}$ and $h_b = Cos^2_{\theta_b}$.
\end{itemize}

Subsequently, the upstream and downstream velocities can also be derived from the aforementioned jump conditions (Eqs. \ref{J1} and \ref{J2}) for a magnetised shock.
\begin{itemize}
 \item SL velocities
 \begin{align}
    v_a = \sqrt{\frac{v_{a1} \pm  \sqrt{h_a} \sqrt{h_b} w_a v_{a2}}{2 \left(h_a w_a+A_s\right) \left(h_a w_a+A_s-h_b w_b\right)}} \nonumber \\
    v_b = \sqrt{\frac{v_{b1} \pm \sqrt{h_a} \sqrt{h_b} w_b v_{b2}}{2 \left(A_s-h_b w_b\right) \left(h_a w_a+A_s-h_b w_b\right)}} \nonumber 
\end{align}
where $w_b= p_b+\epsilon _b,w_a= p_a+\epsilon _a$.
\begin{align*}
    A_s = &p_b - p_a + \frac{B_{by}^2 - B_{ay}^2}{8\pi}\\
    v_{a1} =  &h_a w_a \left(h_b \left(w_a-w_b\right)+2 A_s\right) + 2 A_s \left(A_s-h_b w_b\right) \\
    v_{a2}  =&\sqrt{2 v_{a1} + h_a h_b(w_b^2 - w_a^2)}  \\
    v_{b1} = & 2 A_s \left(h_a w_a-h_b w_b\right)+h_a h_b w_b^2 - h_a w_a h_b w_b+2 A_s^2 \\
    v_{b2}  =& \sqrt{2 v_{b1} + h_a h_b (w_a^2 - w_b^2)}
\end{align*}
\end{itemize}

In the velocity equations, the influence of the magnetic field manifests itself through the magnetic energy density or magnetic pressure, represented by the parameters \( A_s \).

\subsection{The Equation of State}\label{EoS}
We know that to study the effect of shocks in any medium, we need an EoS describing the matter properties.
Since the adiabatic EoS is the most general type of EoS, we adopted such an EoS for this study. Assuming isotropic pressure, the adiabatic gas index, \( \gamma_g \), relates the pressure \( p \) and rest mass density ($\rho$) through the adiabatic expansion law \citep{Timmes_1999ApJS,Henrichs-2000,Baring_2012},
\begin{equation}
   p = K\rho^{\gamma_{g}}.
\end{equation}
In a special relativistic plasma, the adiabatic index \( \gamma_g \) varies from \( \frac{5}{3} \) in the non-relativistic limit to \( \frac{4}{3} \) in the ultra-relativistic regime, reflecting the changing degrees of freedom of the constituent particles \citep{Baring_2012, Aguilar15}. The corresponding internal energy density is given by,
\begin{equation}
    \epsilon = \frac{p}{\gamma_g - 1} + \rho,
\end{equation}
We can express pressure and energy density as a function of number density \citep{Bonazzola_1993, Mallick19},

\begin{align}
 & \epsilon(n) = m_B n + \frac{K\epsilon_0}{\gamma-1}\left(\frac{n}{n_0}\right)^{\gamma_g} \label{24}\\
 & p(n) = {K\epsilon_0}\left(\frac{n}{n_0}\right)^{\gamma_g} \label{25}
\end{align}

where $m_B$ is the baryon mass,  and 

The enthalpy \( w \) can be expressed as,
\begin{equation}
    w = \epsilon + p = \frac{\gamma_g p}{\gamma_g - 1} + \rho.
\end{equation}

While \( \gamma_g \) can be explicitly defined in the asymptotic limits of the non-relativistic and ultra-relativistic regimes, a more rigorous formulation is necessary to accurately describe the transition through the mildly relativistic regime. So, for a relativistic thermal Maxwell-Boltzmann distribution, the Juttner-Synge equation of state provides a more accurate model \citep{Baring_2012},
\begin{equation}
    \frac{w}{\rho} = R(\tau) + \tau, \quad R(\tau) = 3\tau + \frac{K_1(1/\tau)}{K_2(1/\tau)},
\end{equation}
where \( K_1 \) and \( K_2 \) are modified Bessel functions, and \( \tau \) is the dimensionless temperature,
\begin{equation}
    \tau = \frac{kT}{m_e} = \frac{p}{\rho}.
\end{equation}

where $m_e$ is the mass of an electron, $T$ is temperature, and $k$ is Boltzmann constant. To simplify numerical calculations, a Pade approximation for \( R(\tau) \) is used \citep{pade_1994}:
\begin{equation}
    R(\tau) \approx \frac{2 + 7\tau + 12\tau^2 + 6\tau^3}{2 + 4\tau + 2\tau^2}.
\end{equation}

The effective \( \gamma_g \) can then be expressed as,
\begin{equation}
    \gamma_g = 1 + \frac{\tau}{R(\tau) - 1} \approx \frac{5 + 14\tau + 8\tau^2}{3 + 10\tau + 6\tau^2}.
\end{equation}

To incorporate the effects of magnetic fields into our equation of state, we follow the phenomenological approach outlined in \citep{Timmes_1999ApJS, DEXHEIMER_2017, evangelia_2022}. These studies provide a framework for describing the variation of the magnetic field as a function of the chemical potential, which is essential for modelling relativistic shocks and their associated thermodynamic properties in the presence of the magnetic field. The functional form of the magnetic field is given by  

\begin{equation}  
    B(\mu_c) = B_{0} f(\mu_c), \label{B_EoS} 
\end{equation}  

where the scaling function is defined as  

\begin{equation}  
    f(\mu_c) = \frac{a + b\mu_c + c\mu_c^2}{4.419 \times 10^{13}},  
\end{equation}  

with $\mu_c = \frac{p+\epsilon}{n}$, and the constants $a = -0.3$, $b = 0.007$, and $c = 1 \times 10^{-7}$. Here, $B_{0}$ represents the initial magnetic field. Observational studies have reported a range of magnetic field strengths for various stars, with values generally not exceeding $10^4$ Gauss \citep{Henrichs-2000, Donati-2002, Donati-2006, Hubrig-2006, yudin2009stars}. We adopt this upper limit of $B_0 = 10^4$ Gauss in our analysis for simplicity.  

\subsection{The Transport equation}\label{DSA}
An accurate numerical model of DSA requires a well-defined transport equation that describes the evolution of the particle distribution function across the shock front under some EoS. 
The dynamics of the particle distribution function in the context of DSA are described by the following transport equation,
\begin{align}
    &\Gamma \left(1 + v \mu \frac{u}{c^2} \right) \frac{\partial f}{\partial t_s} + \Gamma (v + u \mu) \frac{\partial f}{\partial z_s} = \frac{\partial}{\partial \mu} \left[ D_{\mu \mu}(P) (1 - \mu^2) \frac{\partial f}{\partial \mu} \right],  
    \label{12}
\end{align}  

where \( f = f(p, \mu, z_s, t_s) \) represents the particle distribution function, which depends on the momentum \( p \), pitch angle \( \mu = \cos{\theta} \) (where $\theta$ is the angle between particle's vector and local magnetic field vector), spatial coordinate \( z_s \) (measured along the shock normal), and time \( t_s \). The function \( D_{\mu \mu}(P) \) is the diffusion coefficient, while \( u \) denotes the particle velocity. The Lorentz factor, given by \( \Gamma = \sqrt{1 / (1 - v^2 / c^2)} \), accounts for relativistic corrections \citep{Peacock81,Achterberg01,apeli4}.  

We consider only steady-state solutions in the shock rest frame to simplify the analysis, implying that \( \partial f / \partial t_s = 0 \) \citep{Kirk1987II, KirkAchterberg00}. Under this assumption, the transport equation reduces to a simple form as,  

\begin{equation} \label{13}
    \Gamma (v + u \mu) \frac{\partial f}{\partial z_s} = \frac{\partial}{\partial \mu} \left[ D_{\mu \mu}(P) (1 - \mu^2) \frac{\partial f}{\partial \mu} \right].  
\end{equation}  

This equation describes the evolution of the particle distribution in both the upstream and downstream regions of the shock. 

\subsection{Diffusive Shock Acceleration and Power-Law Energy Spectrum}
DSA is a fundamental process governing the energisation of charged particles in astrophysical shocks. The efficiency of DSA is largely controlled by two key parameters: (i) the average energy gain per shock crossing cycle, which depends on the relative velocities of the upstream and downstream matter, and (ii) the probability that a particle remains confined in the acceleration region rather than escaping downstream, which is determined by the diffusion properties governed by magnetized turbulence and the distribution function of scattering centers. 
%These parameters directly influence the resulting spectral characteristics of accelerated particles and provide crucial insights into the underlying physics of shocks in various astrophysical environments.  

The energy evolution of a single particle undergoing multiple crossings across a shock front follows an exponential growth pattern. If a particle starts with an initial energy \( E_0 \), its energy after a single complete shock crossing cycle is given by  

\begin{equation}
    E = \beta E_0,
\end{equation}  

where \( \beta \) is a constant representing the fractional energy gain per cycle. After \( k \) such cycles, the particle’s energy scales as  

\begin{equation}
    E_k = \beta^k E_0.
    \label{eq:energy_growth}
\end{equation}  

%Since particles must remain in the acceleration region for multiple crossings to achieve high energies, their confinement probability plays a crucial role in shaping the final energy spectrum. 
If \( P_r \) is the probability of a particle remaining within the acceleration region after one complete shock-crossing cycle, the number density of particles still undergoing acceleration after \( k \) cycles follows an exponential decay as,

\begin{equation}
    N_k = P_r^k N_0,
    \label{eq:particle_survival}
\end{equation}  

where \( N_0 \) is the initial number density of injected particles. By expressing \( k \) in terms of the energy gain from Equation \eqref{eq:energy_growth}, the relationship between the number density of particles and their energy can be rewritten as  

\begin{equation}
    \frac{N}{N_0} = \left( \frac{E}{E_0} \right)^{\frac{\ln P_r}{\ln \beta}}.
    \label{eq:number_energy_relation}
\end{equation}  

Thus, the number density of particles decreases as a power-law function of energy. Differentiating Equation \eqref{eq:number_energy_relation} with respect to energy  and noting $\dd N(\geq E) = N(\geq (E + \dd E)) - N(\geq E) =  N(\dd E) = N(E) \dd E$ gives the differential energy distribution as,

\begin{equation}
    N(E) dE \propto E^{-1 + \frac{\ln P_r}{\ln \beta}} dE
    \label{eq:diff_energy_distribution}
\end{equation}  

which can be rewritten as,  

\begin{equation}
    N(E) dE \propto E^{-s} dE
    \label{eq:power_law_spectrum}
\end{equation}  
where \( s \) represents the spectral index of the power-law distribution. Thus, under diffusive shock acceleration, the energy spectrum of accelerated particles naturally follows a power-law dependence, consistent with observational data on cosmic-ray spectra and shock-accelerated particles in astrophysical environments \citep{Ballard91, Achterberg01, Baring_2012}.  

\section{Results and Discussion}\label{sec:5}
The primary objective of this paper is to examine the impact of magnetic fields on shock waves and their influence on the underlying physical processes in DSA. One significant consequence of including the magnetic field is the potential modification of the power-law index in DSA. It is essential to note that the transport equation governing particle acceleration does not explicitly account for magnetic effects. Instead, the influence of the magnetic field enters indirectly through the shock jump conditions and the EoS, which regulate the flow velocities of matter across the shock front. Since the magnetic field alters the velocity structure of the environment, this study aims to quantify its effects on the spectral characteristics of DSA.

%%%%%%%%%%%%%%%%%%%%%%%%%%%%%% Velocities %%%%%%%%%%%%%%%%%%%%%%

\begin{figure*}[ht]
    \centering
\subfloat[]{
  \includegraphics[width=70mm]{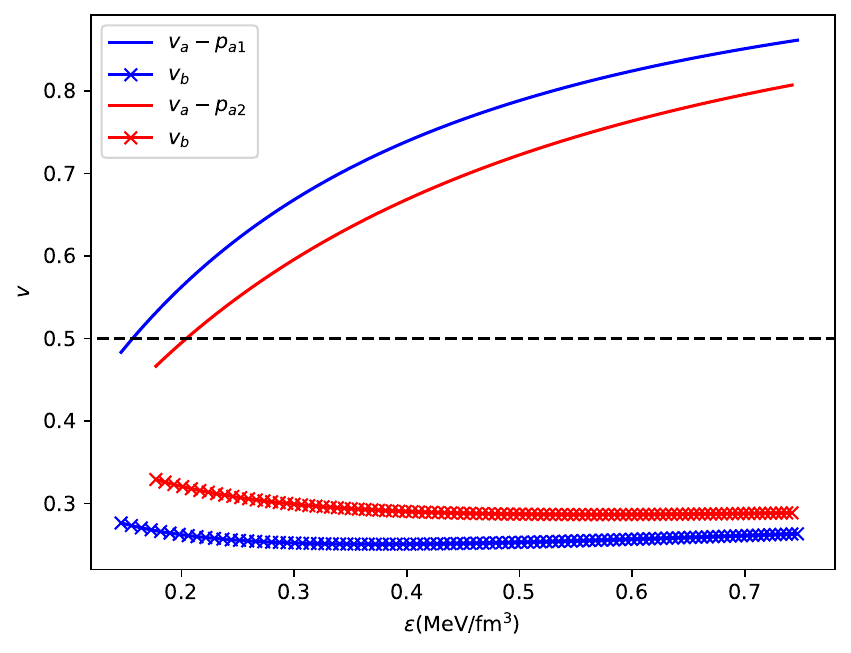}
}
\hspace{0mm}
\subfloat[]{
  \includegraphics[width=70mm]{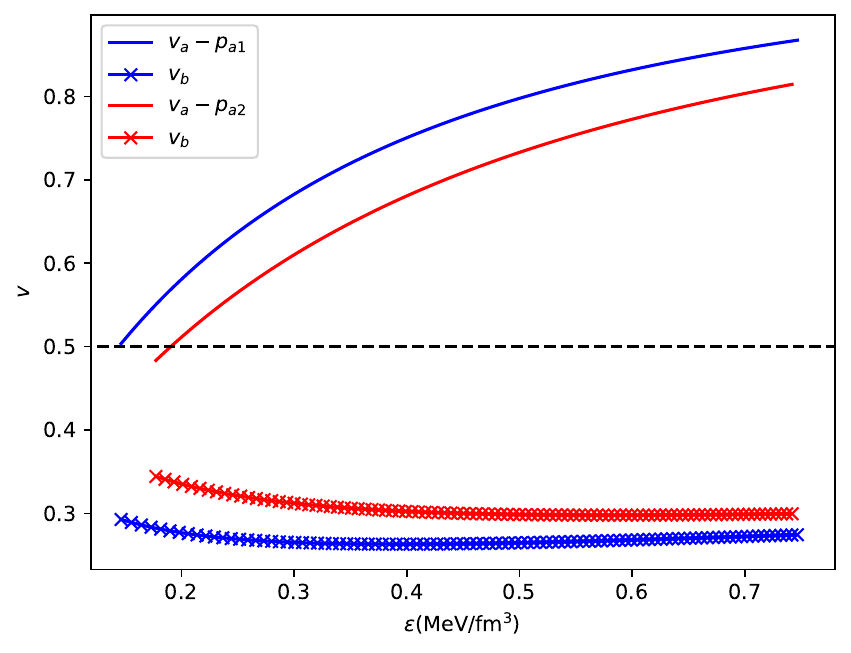}
}
    \caption{The figure illustrates the dependence of upstream and downstream velocities on the energy density (\(\epsilon\)) for an unmagnetized case, with the same analysis in the presence of a magnetic field for comparison. Panel (a) represents the case of an unmagnetized shock, while panel (b) corresponds to a magnetised shock with a zero-incident angle (\(\theta_a \approx 0\)) in both cases. The solid line denotes the upstream velocity, whereas the line with crosses indicates the downstream velocity. The overall trends exhibit similar behaviour across both cases; however, in the unmagnetized scenario, both upstream and downstream velocities are slightly lower for upstream pressures of \( p_{a1} = 0.15 \) MeV/fm\(^3\) and \( p_{a2} = 0.25 \) MeV/fm\(^3\) compared to the magnetized case. The presence of a magnetic field enhances particle scattering and modifies the shock dynamics, leading to marginally higher post-shock velocities. As the upstream pressure increases significantly, the upstream and downstream velocities progressively converge, approaching comparable magnitudes. This trend suggests that the shock undergoes a transition at sufficiently high pressures where magnetic effects become increasingly prominent, reducing the velocity contrast across the discontinuity.}
    \label{f3:v1}
\end{figure*}

\begin{figure*}[ht]
    \centering
\subfloat[]{
  \includegraphics[width=100mm]{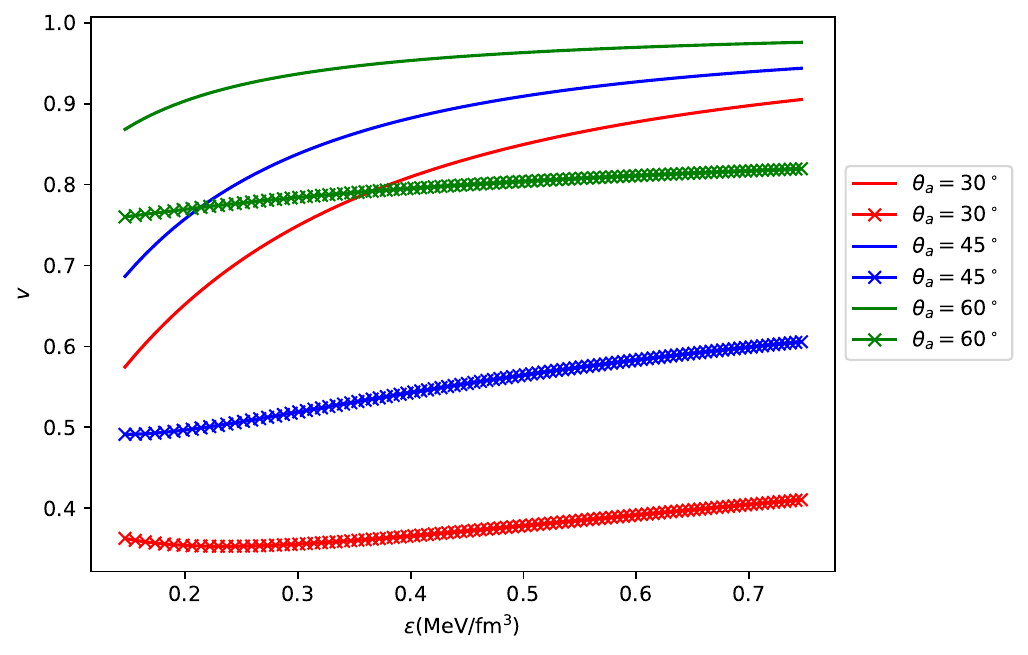}
}
\hspace{0mm}
\subfloat[]{
  \includegraphics[width=100mm]{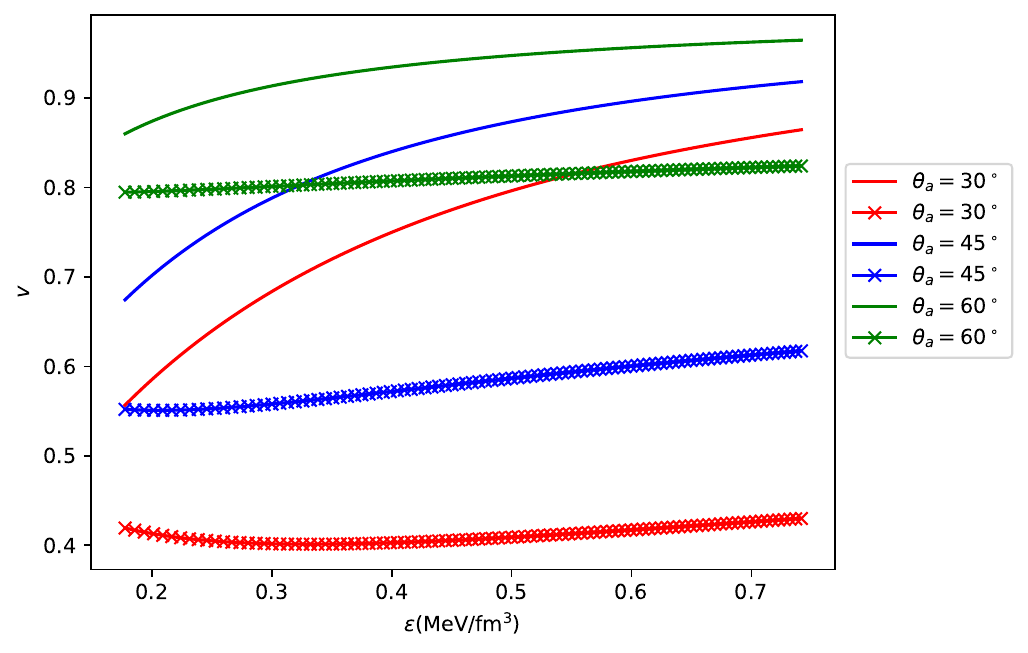}
}
    \caption{The figure illustrates the variation of upstream and downstream velocities as a function of energy density ($\epsilon$) for a magnetised shock. The solid line represents the upstream velocity, while the line marked with crosses depicts the downstream velocity. Figure (a) presents the velocity variations for different incident angles with an upstream pressure of $p_{a1} = 0.15\, MeV/fm^3$, whereas Figure (b) corresponds to an upstream pressure of $p_{a2} = 0.25\, MeV/fm^3$. The downstream pressure can be selected arbitrarily, provided it remains greater than the upstream pressure to ensure the occurrence of a strong shock process. The nature of the plots indicates that as the incident angle increases from lower to higher values, the upstream and downstream velocities progressively converge. This behaviour is consistent with the theoretical expectation that these velocities converge as the incident angle increases.}
    \label{f3:v2}
\end{figure*}

%%%%%%%%%%%%%%%%%%%%%%%%%%%%%%%%%%%%%%%%%%%%%%%%%%%%%%%%%%%%%%%%%%%%%

The results are formulated in the HT frame. Transforming physical quantities from the HT frame to the local fluid rest frame, also referred to as the NI frame, is outlined in Appendix \ref{A1}. Key thermodynamic quantities, such as pressure and number density, remain invariant under this transformation, while velocity and angular variables transform according to a fixed relation. Before analysing the role of magnetisation in DSA, we first characterise the fundamental properties of magnetised shocks and examine how they modify the velocities across the shock front.

\subsection{Study of Magnetised Shock}
To analyse the impact of magnetisation on shock dynamics, we employ the magnetised shock jump conditions to determine the upstream and downstream velocities. 
Since the precise EoS for such astrophysical media remains uncertain, we adopt a representative EoS (described in Section \ref{EoS}) to examine the behaviour of magnetised stellar matter.  Given that the EoS remains uniform throughout the system, we consider two distinct upstream pressure values, which consequently determine the corresponding energy density and magnetic field. Ensuring that the downstream pressure is higher than the upstream pressure allows for the study of the effect of the magnetic field on DSA.

Figures \ref{f3:v1} illustrate the variation of upstream and downstream velocities as a function of energy density for different upstream pressures \( p_{a1} = 0.15\, \text{MeV/fm}^3 \) and \( p_{a2} = 0.25\, \text{MeV/fm}^3 \). Panel (a) represents the case without a magnetic field, while panel (b) includes its effects, which inherently determine the magnetic field strength, both evaluated at zero incident angle. The plots indicate that increasing upstream pressure causes the upstream and downstream velocities to converge, irrespective of the presence of a magnetic field. When the magnetic field is incorporated through the upstream pressure (as described by equation \ref{B_EoS}), both upstream and downstream velocities increase due to the additional contribution from magnetic pressure.

In fig \ref{f3:v1}, the incident angle is kept constant. We plot Fig. \ref{f3:v2} with varying incident angles (\(\theta_a\)). Although a similar trend is observed, a higher magnetic field strength (corresponding to larger values of \(\theta\)) shifts the velocities to higher magnitudes while simultaneously reducing the difference between them. Since the transverse component of the magnetic field on either side of the shock front scales as \( \sin{\theta} \), while the longitudinal component remains unchanged, increasing \(\theta_a\) enhances the overall upstream magnetic pressure. Consequently, this leads to an increase in both upstream and downstream velocities. Moreover, as the upstream pressure increases from \( p_{a1} \) to \( p_{a2} \), the upstream velocity decreases, whereas the downstream velocity exhibits an increasing trend.

%%%%%%%%%%%%%%%%%%%%%%%%%%%%%%%%%%%%%%%%%%%%%%%%%%%%%%%%%%%%%%%%%%%%%%%%%%%%%%%%%%%%%%%%%%%%%%%%%%%%%%%%%%%%%%%%%%%%%%%%%%%%%%%%%%%%%%

\subsection{Application in the Diffusive Shock Acceleration}
Once the aspect of the magnetic field on the shock is analysed, we move on to address our main problem, the magnetic field effect on DSA. DSA occurs when the upstream fluid moves faster relative to the downstream fluid in the shock rest frame, as observed in the study of shock (Fig. \ref{f3:v1} and Fig. \ref{f3:v2}). As particles repeatedly cross the shock, they experience differential motion on both sides, leading to energy gains through Fermi acceleration.

\begin{figure*}[ht]
\centering
\subfloat[]{
  \includegraphics[width=70mm]{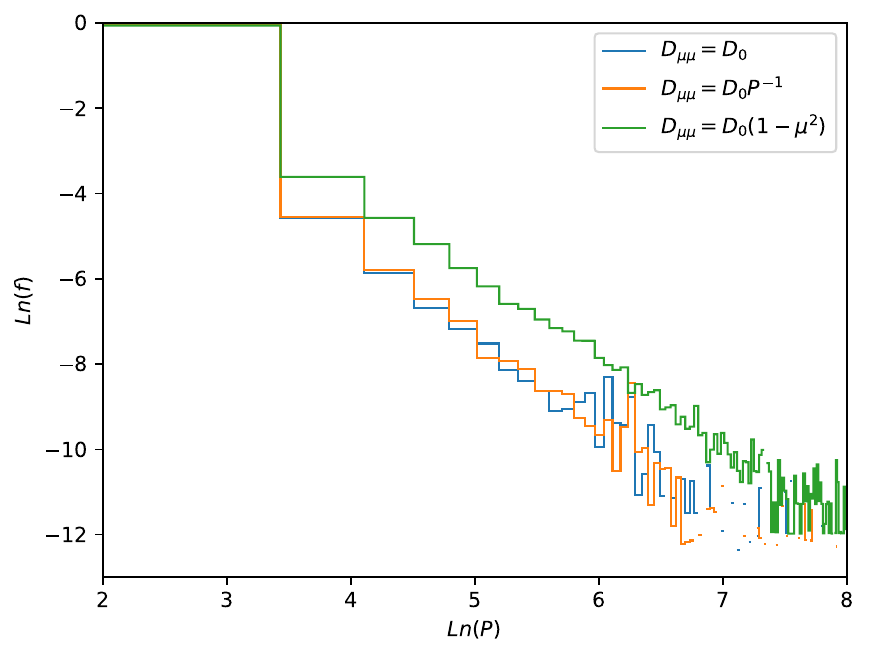}
}
\hspace{0mm}
\subfloat[]{
  \includegraphics[width=70mm]{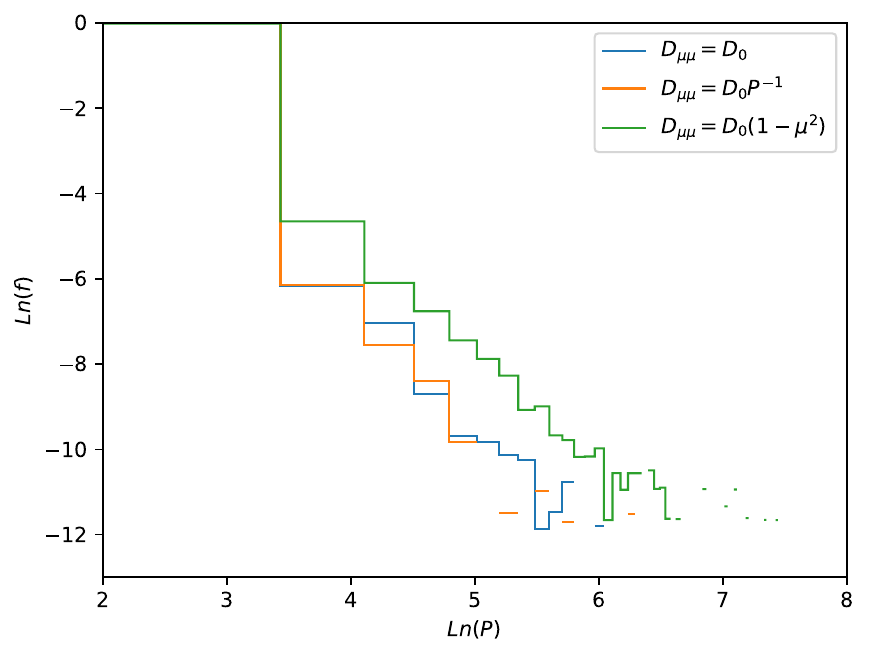}
}
    \caption{Energy spectrum with no magnetic fields under three different diffusion coefficients with two different upstream pressure values: (a) $p_{a1} = 0.15 Mev/fm^3$; (b) $p_{a2} = 0.25 Mev/fm^3$.}
     \label{f5:(without_B)_(50,100)}
\end{figure*}

\begin{figure*}[ht]
    \centering
\subfloat[]{
  \includegraphics[width=70mm]{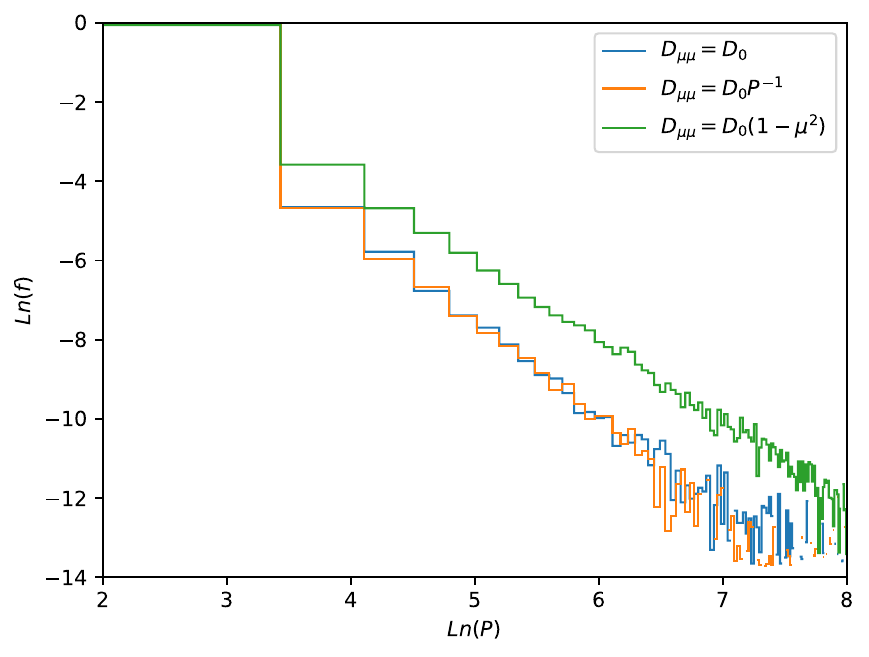}
}
\hspace{0mm}
\subfloat[]{
  \includegraphics[width=70mm]{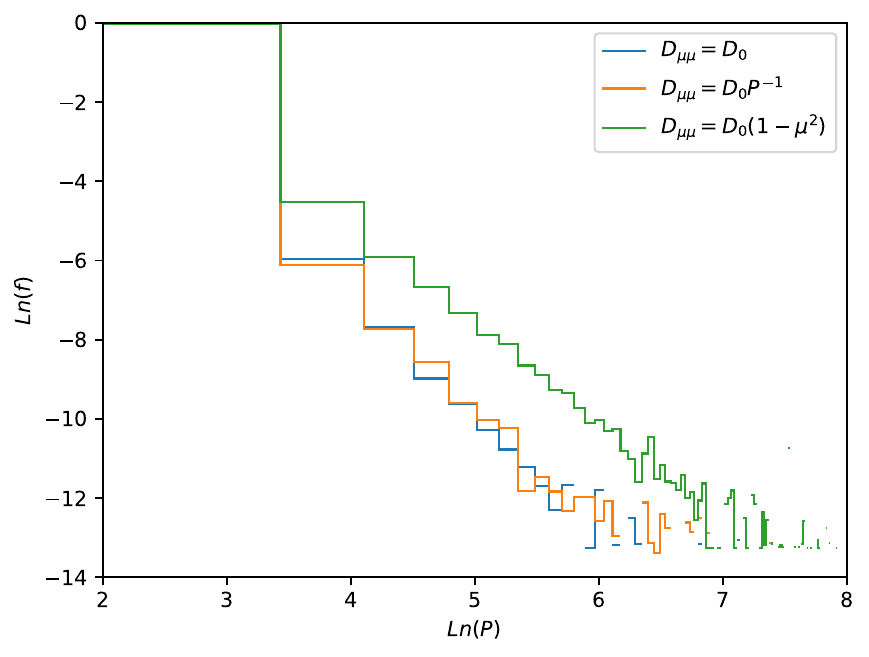}
}
    \caption{The energy spectrum in the presence of a magnetic field at zero incident angle is analyzed for three distinct diffusion coefficients under two different upstream pressure conditions: (a) \( p_{a1} = 0.15 \) MeV/fm\(^3\) and (b) \( p_{a2} = 0.25 \) MeV/fm\(^3\).}
    \label{f5:dsa(B)}
\end{figure*}

%%%%%%%%%%%%%%%%%%%%%%%%%%%%%%%%%%%%%%%%%%%%%%%%%%%%%%%%%%%%%%%%
To study DSA, we employed Monte Carlo simulations based on methodologies established in the literature \citep{Ellison83, Kirk1987I, Kirk1987II, KirkAchterberg00}. By simulating a large number of particles over an adequate number of steps, we observed the emergence of a power-law momentum distribution. The spectral index, \( s \), was extracted by fitting a linear regression to the logarithmic plot of the distribution function \( f \) (number of particles distribution) versus momentum \( P \) (measure of energy), where the slope corresponds to \( -s \). The input parameters for the simulation include the upstream and downstream velocities (\( v_a \) and \( v_b \)) measured in the shock rest frame, which defines the nature of the shock under consideration \citep{Kirk99, Achterberg01,arbutina_2021}.

\begin{figure*}[ht]
    \centering
\subfloat[]{
  \includegraphics[width=70mm]{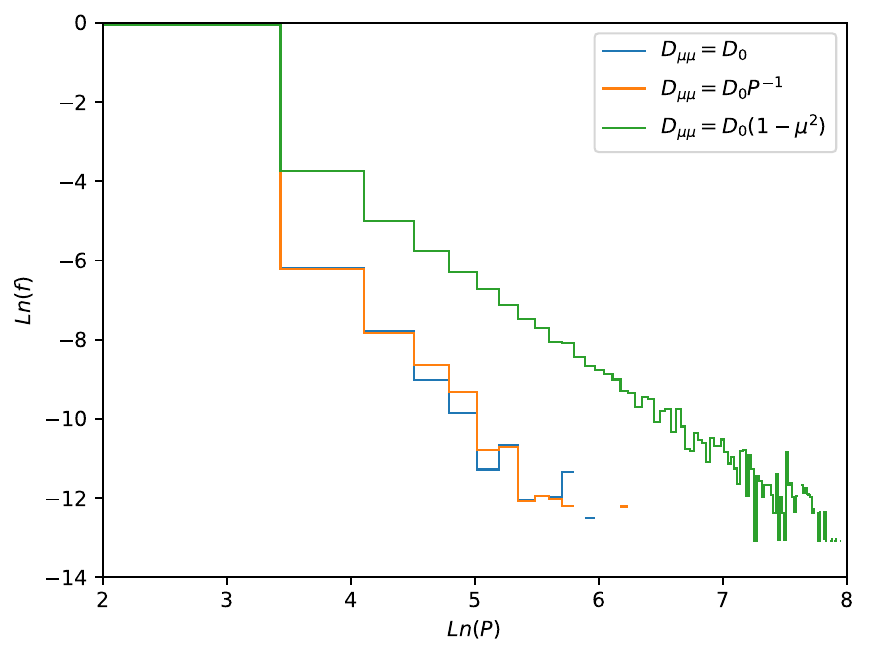}
}
\hspace{0mm}
\subfloat[]{
  \includegraphics[width=70mm]{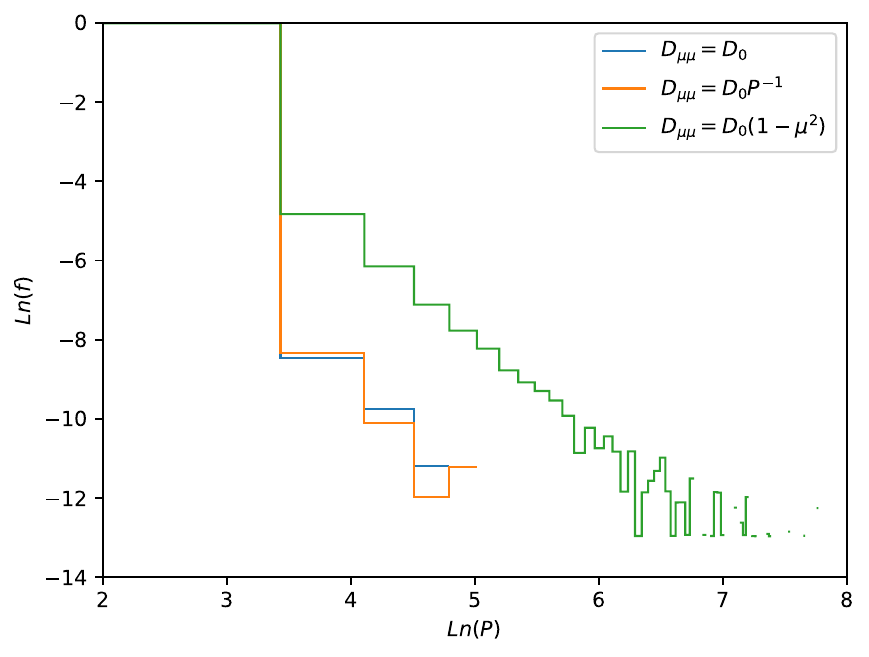}
}
    \caption{The energy spectra for magnetised shocks are presented, derived using Monte Carlo simulations based on diffusive shock acceleration. The x-axis denotes the logarithm of particle momentum, while the y-axis represents the logarithm of the particle distribution function. DSA Spectrum of variable upstream and downstream velocities with $\theta_{a}=30^\circ$. Each plot represents 3 cases of constant diffusion coefficient, momentum and pitch angle dependence. Plots correspond to two different upstream pressure values, (a) $p_{a1} = 0.15 MeV/fm^3$; (b) $p_{a2} = 0.25 MeV/fm^3$. }
    \label{f5:(D_constant)_30_(50,100)}
\end{figure*}

Our simulations explored different forms of the diffusion coefficient, \( D_{\mu \mu}(P) \), depending on the power spectrum of magnetic fluctuations. Under quasi-linear theory, where the power spectrum is inversely proportional to the wavenumber \( k \), the diffusion coefficient remains independent of pitch angle \( \mu \) and momentum, taking the form \( D_{\mu \mu}(P) = D_{0} \) in the absence of radiation losses \citep{Kirk1987I}. However, for a Kolmogorov-type spectrum, \( D_{\mu \mu} \) exhibits a dependence on pitch angle and is modeled as \( D_{\mu \mu}(P)= D_0 (1 - \mu^2) \) \citep{Kirk99}. Additionally, considering the Fokker-Planck transport equation, the spatial distribution of particles near shocks is significantly influenced by the momentum dependence of \( D_{\mu \mu}(P) \). To account for this, we assume a power-law dependence, adopting the form \( D_{\mu \mu}(P) = D_0 P^{-1} \), where \( D_0 \) represents the isotropic diffusion coefficient \citep{Kirk1987II}.  

\begin{figure*}[ht]
\centering
{
  \includegraphics[width=70mm]{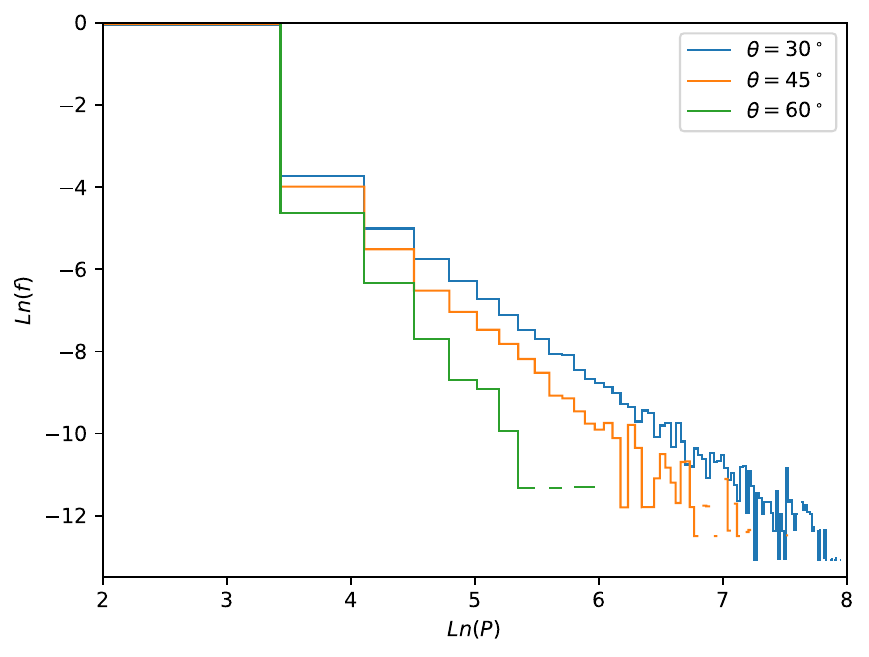}
}
% \hspace{0mm}
% \subfloat[]{
%   \includegraphics[width=80mm]{used_results/withB_muchange_100.pdf}
% }
    \caption{DSA spectra for varying upstream and downstream velocities at different incident angles, computed for an upstream pressure of $p_{a1} = 0.15 \, \text{MeV/fm}^3$. The plot includes three cases of incident angles, with a pitch angle-dependent diffusion coefficient.}
     \label{f5:(mu_change)_(50,100)}
\end{figure*}

To ensure the reliability of our simulation framework, we first validated our approach by determining the spectral index for non-relativistic diffusive shock acceleration (DSA), following the established methodologies of \cite{Ellison83, Kirk1987I, Kirk1987II}. Once validated, the analysis was extended to relativistic shocks, considering cases both with and without magnetic fields. The simulations were carried out with a statistically significant ensemble of particles (\(N=20,000 \)) to ensure robustness in spectral index determination. The spectral indices were extracted by performing least-squares fitting on the log-log representation of particle distribution function (\( f \)) versus momentum (\( P \)), confirming the expected power-law behaviour across different shock scenarios.  

In order to assess the influence of magnetic fields on energy spectra, we begin by examining the spectral characteristics in a non-magnetised medium. The results, depicted in Figures \ref{f5:(without_B)_(50,100)}a and \ref{f5:(without_B)_(50,100)}b, demonstrate that the spectral index is strongly dependent on the chosen diffusion mechanism. For both constant and momentum-dependent diffusion coefficients, the spectral indices are found within the range \( s \approx 2.5 - 2.6 \) in Figure \ref{f5:(without_B)_(50,100)}a (corresponding to both the values of $p_{a1}$) and \( s \approx 3.1 - 3.2 \) (corresponding to both the values of $p_{a1}$) in Figure \ref{f5:(without_B)_(50,100)}b. In contrast, when diffusion is governed by pitch-angle dependence, the resulting spectral indices are systematically lower, approximately \( s \approx 2.05 - 2.10 \) in Figure \ref{f5:(without_B)_(50,100)}a and \( s \approx 2.2 - 2.3 \) in Figure \ref{f5:(without_B)_(50,100)}b. Furthermore, increasing the upstream pressure leads to a systematic steepening of the spectra across all diffusion models. However, the pitch-angle-dependent diffusion mechanism exhibits a weaker response to this pressure increase compared to the other two cases. This behaviour suggests that as pressure increases, particle acceleration results in higher momentum, leading to steeper spectra for constant and momentum-dependent diffusion coefficients. In contrast, the pitch-angle-dependent diffusion mechanism enhances particle confinement, reducing the direct impact of increased momentum on spectral steepening.

The presence of a magnetic field significantly modifies the spectral characteristics, leading to an overall flattening of the energy spectrum across all diffusion models. This indicates a more efficient acceleration process due to enhanced particle retention within the acceleration region. Specifically, for the constant and momentum-dependent diffusion coefficients, the spectral indices reduce to \( s \approx 2.2 - 2.3 \) at an upstream pressure of \( p_{a1} = 0.15 \) MeV/fm\(^3\) (Figure \ref{f5:dsa(B)}a) and \( s \approx 2.7 - 2.8 \) at \( p_{a2} = 0.25 \) MeV/fm\(^3\) (Figure \ref{f5:dsa(B)}b). A similar spectral flattening is observed for the pitch-angle-dependent diffusion coefficient, with indices shifting to \( s \approx 1.9 - 2.0 \) in Figure \ref{f5:dsa(B)}a and \( s \approx 2.1 - 2.2 \) in Figure \ref{f5:dsa(B)}b.  

In a non-magnetised environment, particle transport is predominantly regulated by turbulence-induced scattering, which is insufficient to confine particles effectively within the acceleration zone. As a result, particles experience rapid escape, leading to steep spectra. The inefficiency of confinement in such scenarios significantly limits the number of shock crossings, thereby reducing the maximum energy that particles can attain. However, when a magnetic field is introduced, the situation changes dramatically. The increased level of magnetic turbulence enhances the scattering efficiency, allowing particles to undergo multiple shock crossings before escaping. This extended residence time leads to more effective acceleration and consequently produces a flatter spectral slope. Notably, pitch-angle-dependent diffusion appears to be the most sensitive to the presence of a magnetic field, yielding the flattest energy spectra, which underscores its strong role in particle retention and acceleration within magnetised shocks.

Further insight into magnetisation effects on spectral characteristics is gained by examining the influence of an oblique upstream magnetic field at an inclination angle of \(30^\circ\). The corresponding energy spectra, illustrated in Figures \ref{f5:(D_constant)_30_(50,100)}a and \ref{f5:(D_constant)_30_(50,100)}b, are obtained for upstream pressures of \( p_{a1} = 0.15 \) MeV/fm\(^3\) and \( p_{a2} = 0.25 \) MeV/fm\(^3\), respectively. Under these conditions, the shock remains in a strong shock regime, where the downstream pressure significantly exceeds the upstream pressure \citep{Landau87,ansh2}. The analysis of energy spectra under three distinct diffusion regimes, constant, momentum-dependent, and pitch-angle-dependent, reveals that both constant and momentum-dependent diffusion coefficients yield steeper spectral indices within the range \( s \approx 3.2 - 3.4 \) (smaller value corresponds to smaller $p_{a1}$). However, the pitch-angle-dependent diffusion coefficient produces a systematically flatter spectrum, with spectral indices in the range \( s \approx 2.3 - 2.4 \). This discrepancy arises because pitch-angle-dependent diffusion enables a higher frequency of particle re-crossings across the shock interface, effectively prolonging their confinement in the acceleration region. The extended particle residence time facilitates additional energy gain, leading to a comparatively shallower spectral slope than those produced by the other diffusion models.

An additional factor influencing spectral formation is the interplay between shock inclination angle and velocity. As the upstream magnetic field inclination increases, the downstream velocity approaches the upstream velocity. This trend directly impacts the spatial extent of the acceleration zone, as a higher downstream velocity reduces the confinement region, making it increasingly challenging for an energy spectrum to form. Despite these variations, pitch-angle-dependent diffusion remains particularly effective in sustaining a well-defined energy spectrum, regardless of changes in upstream/downstream velocity ratios or pressure conditions. Its efficiency in maintaining prolonged particle residence times ensures continued acceleration and spectral robustness across diverse shock configurations.

A more detailed examination of the pitch-angle-dependent diffusion coefficient, as depicted in Figure \ref{f5:(mu_change)_(50,100)}, confirms that the energy spectrum remains stable across different incident angles. This suggests that even as the downstream velocity increases, the pitch-angle-dependent diffusion mechanism compensates by extending the confinement time of particles, thereby preserving spectral integrity across varying upstream conditions. Furthermore, Figure \ref{f5:(mu_change)_(50,100)} indicates that the spectral slope systematically increases with the upstream incidence angle \( \theta_a \), ranging from approximately 2.1 to 2.4. Additionally, an increase in upstream pressure leads to further modifications in the spectral characteristics, following the same trend observed previously. A higher upstream pressure is associated with greater magnetic turbulence, which, in turn, enhances the efficiency of particle scattering and results in a steeper spectral index.

Without magnetisation, diffusion mechanisms struggle to maintain extended particle confinement, resulting in steeper spectra. However, in the presence of a magnetic field, enhanced turbulence facilitates stronger scattering, leading to more efficient particle acceleration and ultimately producing flatter energy spectra. Among the diffusion models studied, pitch-angle-dependent diffusion exhibits the strongest response to magnetisation, demonstrating its superior ability to sustain prolonged acceleration and spectral flattening in magnetised shock environments.

\section{Summary and Conclusion}\label{sec:6}
The study systematically examines the impact of the magnetic field on shock dynamics and its subsequent effects on DSA. The magnetic field effect is incorporated into the EoS and the energy-momentum tensor, and it is reflected in the shock jump conditions. It ultimately affects the upstream and downstream velocities, which are incorporated into the transport equation within a Monte Carlo simulation framework. Our results confirm that in magnetised shocks, the interplay between the EoS and shock magnetisation significantly influences the final energy distribution of accelerated particles. The angular dependence introduced by magnetisation leads to the convergence of upstream and downstream velocities at higher incidence angles, affecting the spatial extent of the acceleration region and consequently shaping the resulting energy spectrum. Our results demonstrate that, in the absence of a magnetic field, energy spectra tend to be steeper due to inefficient particle confinement and rapid escape from the acceleration region. The spectral indices vary significantly across different diffusion models, with pitch-angle-dependent diffusion consistently yielding lower spectral slopes compared to constant and momentum-dependent diffusion. This suggests that pitch-angle-dependent diffusion facilitates extended particle confinement, even in non-magnetised shocks, although the overall acceleration efficiency remains limited under such conditions.

Introducing a magnetic field fundamentally alters the acceleration process by enhancing particle scattering and prolonging confinement within the acceleration region. Across all diffusion models, the energy spectra become systematically flatter, indicating more efficient energy gain through repeated shock crossings. The presence of a magnetic field strengthens turbulence, allowing for enhanced pitch-angle scattering, particularly in the pitch-angle-dependent diffusion regime. Consequently, this diffusion model exhibits the strongest flattening effect, demonstrating that magnetised shocks provide a more favourable environment for particle acceleration.

Furthermore, we investigated the role of upstream pressure variations in shaping spectral properties. A higher upstream pressure, associated with stronger magnetic turbulence, leads to a broader spectral range, with both flatter and steeper slopes observed at different energy scales. This is attributed to increased particle scattering efficiency, which extends the duration of shock crossings and influences the energy gain process. Additionally, as the shock inclination angle increases, the velocity contrast between upstream and downstream regions diminishes, impacting the extent of the acceleration zone. Nevertheless, the pitch-angle-dependent diffusion mechanism remains effective across all scenarios, ensuring sustained particle acceleration and a well-defined energy spectrum under varying shock conditions. 

Overall, our findings reinforce the importance of magnetic field effects in shaping the energy spectra of accelerated particles. While non-magnetised shocks exhibit steeper spectral slopes due to rapid particle escape, magnetised shocks significantly enhance confinement, leading to flatter spectra across all diffusion regimes. The pitch-angle-dependent diffusion coefficient remains the most efficient in sustaining a well-defined spectrum under varying shock conditions, emphasising its role in facilitating high-energy particle acceleration in astrophysical shocks.  

The Monte Carlo simulation developed for modelling DSA in both fast and slow shocks with specific diffusion characteristics has broad astrophysical applications. By incorporating relativistic EoS, we obtained more precise estimates of ambient fluid properties, improving the ability to identify shock types responsible for observed cosmic ray spectra. At present, the diffusion equation assumes no non-linear dependence on the distribution function \citep{arbutina_2021, LEVAN201544}. Future work may explore a more realistic approach, incorporating non-linear terms, particularly to account for stochastic scattering by Alfvén waves influenced by the particle distribution function. Further studies should incorporate these effects, as well as general relativistic treatment \citep{ansh1,ansh2}, to further improve spectral predictions.

\section*{Acknowledgements}

The authors thank the Indian Institute of Science Education and Research, Bhopal, for providing all the research and infrastructure facilities. AV acknowledges the Prime Minister’s Research Fellowship (PMRF), Ministry of Education, Government of India, for a graduate fellowship, and RM acknowledges the Science and Engineering Research Board (SERB), Government of India, for monetary support in the form of a Core Research Grant (CRG/2022/000663).

\bibliographystyle{JCAP}

% Loading bibliography database
\bibliography{bibliography}

@article{Fermi49,
  title = {On the Origin of the Cosmic Radiation},
  author = {Fermi, ENRICO},
  journal = {Phys. Rev.},
  volume = {75},
  issue = {8},
  pages = {1169--1174},
  numpages = {0},
  year = {1949},
  month = {4},
  publisher = {American Physical Society},
  doi = {10.1103/PhysRev.75.1169},
  url = {https://link.aps.org/doi/10.1103/PhysRev.75.1169}
}

@article{Aflven43,
       author = {{Alfv{\'e}n}, H.},
        title = "{On the Existence of Electromagnetic-Hydrodynamic Waves}",
      journal = {Arkiv for Matematik, Astronomi och Fysik},
         year = 1943,
        month = jan,
       volume = {29B},
        pages = {1-7},
       adsurl = {https://ui.adsabs.harvard.edu/abs/1943ArMAF..29B...1A}
}

@article{Bell78,
    author = {Bell, A. R.},
    title = "{The acceleration of cosmic rays in shock fronts – I}",
    journal = {Monthly Notices of the Royal Astronomical Society},
    volume = {182},
    number = {2},
    pages = {147-156},
    year = {1978},
    month = {02},
    issn = {0035-8711},
    doi = {10.1093/mnras/182.2.147},
    url = {https://doi.org/10.1093/mnras/182.2.147}
}

@article{BlandfordOstriker78,
    author = {{Blandford}, R.D. and {Ostriker}, J.P.},
    title = {Particle acceleration by astrophysical shocks.},
      journal = {The Astrophysical Journal Letters},
      year = {1978},
      month = {04},
      volume = {221},
      pages = {L29-L32},
      doi = {10.1086/182658},
      adsurl = {https://ui.adsabs.harvard.edu/abs/1978ApJ...221L..29B}
}

@book{Landau87,
  author = {Landau, L. D. and Lifshitz, E. M.},
  edition = 2,
  isbn = {9780080570730},
  month = 1,
  publisher = {Pergamon Press},
  series = {Course of theoretical physics / by L. D. Landau and E. M. Lifshitz, Vol. 6},
  title = {Fluid Mechanics, Second Edition: Volume 6 (Course of Theoretical Physics)},
  url = {https://www.elsevier.com/books/fluid-mechanics/landau/978-0-08-057073-0},
  year = 1987
}

@book{Csernai94,
    author = "Csernai, L. P.",
    publisher = {John Wiley \& Sons Ltd},
    title = "{Introduction to relativistic heavy ion collisions}",
    year = "1994",
    url = {http://www.csernai.no/Csernai-textbook.pdf}
}

@article{Blandford87,
    author = {{Blandford}, Roger and {Eichler}, David},
    title = "{Particle acceleration at astrophysical shocks: A theory of cosmic ray origin.}",
    journal = {Physics Reports}, 
    year = 1987,
    volume = {154},
    number = {1},
    pages = {1-75},
    doi = {https://doi.org/10.1016/0370-1573(87)90134-7},
    issn = {0370-1573},
    url = {https://www.sciencedirect.com/science/article/pii/0370157387901347}
    }

@book{Longair11,
  title     = "{High Energy Astrophysics (Third Edition)}",
  author    = "M. Longair",
  year      = 2011,
  publisher = "Cambridge University Press",
  address   = "Cambridge, UK"
}

@misc{Gallant02,
      title="{Particle Acceleration at Relativistic Shocks}", 
      author={Yves A. Gallant},
      year={2002},
      eprint={astro-ph/0201243},
      archivePrefix={arXiv},
      primaryClass={astro-ph}
}

@book{Alan89,
    title = "{Computer Simulation of Liquids}",
    author = {Allen, M. P. and Tildesley, D. J.},
    year = {1989},
    isbn = {0198556454},
    publisher = {Clarendon Press},
    address = {Oxford}
}

@article{Krymskii77,
       author = {{Krymskii}, G.~F.},
        title = "{A regular mechanism for the acceleration of charged particles on the front of a shock wave}",
      journal = {Akademiia Nauk SSSR Doklady},
         year = 1977,
        month = jun,
       volume = {234},
        pages = {1306-1308},
       adsurl = {https://ui.adsabs.harvard.edu/abs/1977DoSSR.234.1306K}
}

@inproceedings{Axford77,
       author = {{Axford}, W.~I. and {Leer}, E. and {Skadron}, G.},
        title = "{The Acceleration of Cosmic Rays by Shock Waves}",
    booktitle = {International Cosmic Ray Conference},
         year = 1977,
       series = {International Cosmic Ray Conference},
       volume = {11},
        month = jan,
        pages = {132},
       adsurl = {https://ui.adsabs.harvard.edu/abs/1977ICRC...11..132A}
}

@inproceedings{Ellison83,
       author = {{Ellison}, D.~C. and {Jones}, F.~C. and {Eichler}, D.},
       title = "{Monte Carlo Simulation of Steady State Shock Structure Including Cosmic Ray Mediation and Particle Escape}",
       booktitle = {International Cosmic Ray Conference},
         year = 1983,
       series = {International Cosmic Ray Conference},
       volume = {2},
        month = aug,
        pages = {271},
       adsurl = {https://ui.adsabs.harvard.edu/abs/1983ICRC....2..271E}
}

@article{Achterberg01,
	doi = {10.1046/j.1365-8711.2001.04851.x},
  
	url = {https://doi.org/10.1046\%2Fj.1365-8711.2001.04851.x},
  
	year = 2001,
	month = {12},
  
	publisher = {Oxford University Press ({OUP})},
  
	volume = {328},
  
	number = {2},
  
	pages = {393--408},
  
	author = {A. Achterberg and Y. A. Gallant and J. G. Kirk and A. W. Guthmann},
  
	title = {Particle acceleration by ultrarelativistic shocks: theory and simulations},
  
	journal = {Monthly Notices of the Royal Astronomical Society}
}

@article{Peacock81,
    author = {Peacock, J. A.},
    title = "{Fermi acceleration by relativistic shock waves}",
    journal = {Monthly Notices of the Royal Astronomical Society},
    volume = {196},
    number = {2},
    pages = {135-152},
    year = {1981},
    month = {09},
    issn = {0035-8711},
    doi = {10.1093/mnras/196.2.135},
    url = {https://doi.org/10.1093/mnras/196.2.135}
}

@article{Gaisser2006,
    author = {T. K. Gaisser},
    title = "{The Cosmic-ray Spectrum: from the knee to the ankle}",
    journal = {Journal of Physics: Conference Series},
    volume = {47},
    pages = {15-20},
    year = {2006},
    issn = {1742-6596; 1742-6588},
    doi = {10.1088/1742-6596/47/1/002},
    url = {https://dx.doi.org/10.1088/1742-6596/47/1/002}
}

@article{Kirk1987I,
    author = {J. G. Kirk and P. Schneider},
    title = "{On the Acceleration of Charged Particles at Relativistic Shock Fronts}",
    journal = {The Astrophysical Journal},
    volume = {315},
    pages = {425-433},
    year = {1987},
    month = {04},
    issn = {0004-637X},
    doi = {10.1086/165147},
    url = {https://adsabs.harvard.edu/full/1987ApJ...315..425K},
}

@article{Kirk1987II,
    author = {J. G. Kirk and P. Schneider},
    title = "{Particle Acceleration at Shocks: A Monte Carlo Method}",
    journal = {The Astrophysical Journal},
    volume = {322},
    pages = {256-265},
    year = {1987},
    month = {11},
    issn = {0004-637X},
    doi = {10.1086/165720},
    url = {https://adsabs.harvard.edu/full/1987ApJ...322..256K},
}

@article{KirkAchterberg00,
    author = {Achterberg, Abraham and Gallant, Yves A. and Kirk, John G. and Guthmann, Axel W.},
    title = "{Particle acceleration by ultrarelativistic shocks: theory and simulations}",
    journal = {Monthly Notices of the Royal Astronomical Society},
    volume = {328},
    number = {2},
    pages = {393-408},
    year = {2001},
    month = {12},
    issn = {0035-8711},
    doi = {10.1046/j.1365-8711.2001.04851.x},
    url = {https://doi.org/10.1046/j.1365-8711.2001.04851.x}
}

@article{Ballard91,
    author = {K. R. Ballard and A. F. Heavens},
    title = "{First-order Fermi acceleration at oblique relativistic magnetohydrodynamic shocks}",
    journal = {Monthly Notices of the Royal Astronomical Society},
    volume = {251},
    pages = {438-448},
    year = {1991},
    month = {08},
    issn = {0035-8711},
    doi = {10.1093/mnras/251.3.438},
    url = {https://doi.org/10.1093/mnras/251.3.438}
}

@article{Mallick19,
	doi = {10.1093/mnras/stz454},
  
	url = {https://doi.org/10.1093%2Fmnras%2Fstz454},
  
	year = 2019,
	month = {2},
  
	publisher = {Oxford University Press ({OUP})},
  
	volume = {485},
  
	number = {1},
  
	pages = {577--585},
  
	author = {Ritam Mallick and Mohammad Irfan},
  
	title = {Combustion adiabat and the maximum mass of a quark star},
  
	journal = {Monthly Notices of the Royal Astronomical Society}
}

@ARTICLE{Arzner04,
       author = {{Arzner}, Kaspar and {Vlahos}, Loukas},
        title = "{Particle Acceleration in Multiple Dissipation Regions}",
      journal = {The Astrophysical Journall},
         year = 2004,
        month = apr,
       volume = {605},
       number = {1},
        pages = {L69-L72},
          doi = {10.1086/392506},
archivePrefix = {arXiv},
       eprint = {astro-ph/0402605},
 primaryClass = {astro-ph},
       adsurl = {https://ui.adsabs.harvard.edu/abs/2004ApJ...605L..69A}
}

@ARTICLE{Arzner06,
       author = {{Arzner}, Kaspar and {Knaepen}, Bernard and {Carati}, Daniele and {Denewet}, Nicolas and {Vlahos}, Loukas},
        title = "{The Effect of Coherent Structures on Stochastic Acceleration in MHD Turbulence}",
      journal = {The Astrophysical Journal},
         year = 2006,
        month = jan,
       volume = {637},
       number = {1},
        pages = {322-332},
          doi = {10.1086/498341},
archivePrefix = {arXiv},
       eprint = {astro-ph/0509717},
 primaryClass = {astro-ph},
       adsurl = {https://ui.adsabs.harvard.edu/abs/2006ApJ...637..322A}
}

@BOOK{fokker89,
       author = {{Risken}, H.},
       publisher = "Springer Berlin, Heidelberg",
        title = "{The Fokker-Planck equation. Methods of solution and applications}",
         year = 1989,
       adsurl = {https://ui.adsabs.harvard.edu/abs/1989fpem.book.....R}
}

@article{Bell04,
    author = {Bell, A. R.},
    title = "{Turbulent amplification of magnetic field and diffusive shock acceleration of cosmic rays}",
    journal = {Monthly Notices of the Royal Astronomical Society},
    volume = {353},
    number = {2},
    pages = {550-558},
    year = {2004},
    month = {09},
    issn = {0035-8711},
    doi = {10.1111/j.1365-2966.2004.08097.x},
    url = {https://doi.org/10.1111/j.1365-2966.2004.08097.x}
}

@ARTICLE{kirkdendy01,
       author = {{Kirk}, J.~G. and {Dendy}, R.~O.},
        title = "{Shock acceleration of cosmic rays - a critical review}",
      journal = {Journal of Physics G Nuclear Physics},
         year = 2001,
        month = jul,
       volume = {27},
       number = {7},
        pages = {1589-1595},
          doi = {10.1088/0954-3899/27/7/316},
archivePrefix = {arXiv},
       eprint = {astro-ph/0101175},
 primaryClass = {astro-ph},
       adsurl = {https://ui.adsabs.harvard.edu/abs/2001JPhG...27.1589K}
}

@ARTICLE{Drury83,
       author = {{Drury}, L. Oc.},
        title = "{REVIEW ARTICLE: An introduction to the theory of diffusive shock acceleration of energetic particles in tenuous plasmas}",
      journal = {Reports on Progress in Physics},
         year = 1983,
        month = aug,
       volume = {46},
       number = {8},
        pages = {973-1027},
          doi = {10.1088/0034-4885/46/8/002},
       adsurl = {https://ui.adsabs.harvard.edu/abs/1983RPPh...46..973D}
}

@ARTICLE{Berezhko02,
       author = {{Berezhko}, E.~G. and {Ksenofontov}, L.~T. and {V{\"o}lk}, H.~J.},
        title = "{Emission of SN 1006 produced by accelerated cosmic rays}",
      journal = {Astronomy \& Astrophysics (A\&A)},
         year = 2002,
        month = Dec,
       volume = {395},
        pages = {943-953},
          doi = {10.1051/0004-6361:20021219},
archivePrefix = {arXiv},
       eprint = {astro-ph/0204085},
 primaryClass = {astro-ph},
       adsurl = {https://ui.adsabs.harvard.edu/abs/2002A\&A...395..943B}
}

@ARTICLE{Prishchep81,
       author = {{Prishchep}, V.~L. and {Ptuskin}, V.~S.},
        title = "{The acceleration of fast particles at the front of a spherical shock wave}",
      journal = {The Astrophysical Journal },
         year = 1981,
        month = aug,
       volume = {58},
        pages = {779-789},
       adsurl = {https://ui.adsabs.harvard.edu/abs/1981AZh....58..779P}
}

@article{Neronov17,
	author = {{Neronov, Andrii} and {Malyshev, Denys} and {Semikoz, Dmitri V.}},
	title = {Cosmic-ray spectrum in the local Galaxy},
	DOI= "10.1051/0004-6361/201731149",
	url= "https://doi.org/10.1051/0004-6361/201731149",
	journal = {A\&A},
	year = 2017,
	volume = 606,
	pages = "A22",
}

@article{neronov15,
  title={Hard spectrum of cosmic rays in the Disks of Milky Way and Large Magellanic Cloud},
  author={Neronov, A and Malyshev, D},
  journal={arXiv preprint arXiv:1505.07601},
  year={2015}
}

@ARTICLE{Yang14,
       author = {{Yang}, Rui-zhi and {de O{\~n}a Wilhelmi}, Emma and {Aharonian}, Felix},
        title = "{Probing cosmic rays in nearby giant molecular clouds with the Fermi Large Area Telescope}",
      journal = {Astronomy \& Astrophysics (A\&A)},
         year = 2014,
        month = jun,
       volume = {566},
          eid = {A142},
        pages = {A142},
          doi = {10.1051/0004-6361/201321044},
archivePrefix = {arXiv},
       eprint = {1303.7323},
 primaryClass = {astro-ph.HE},
       adsurl = {https://ui.adsabs.harvard.edu/abs/2014A\& A...566A.142Y}
}

@article{acero15,
  title={Fermi large area telescope third source catalog},
  author={Acero, Fabio and Ackermann, Markus and Ajello, M and Albert, A and Atwood, WB and Axelsson, Magnus and Baldini, Luca and Ballet, J and Barbiellini, G and Bastieri, Denis and others},
  journal={The Astrophysical Journal Supplement Series},
  volume={218},
  number={2},
  pages={23},
  year={2015},
  publisher={IOP Publishing}
}

@ARTICLE{Adriani14,
       author = {{Adriani}, O. and {Barbarino}, G.~C. and {Bazilevskaya}, G.~A. and {Bellotti}, R. and {Boezio}, M. and {Bogomolov}, E.~A. and {Bongi}, M. and {Bonvicini}, V. and {Bottai}, S. and {Bruno}, A. and {Cafagna}, F. and {Campana}, D. and {Carbone}, R. and {Carlson}, P. and {Casolino}, M. and {Castellini}, G. and {Danilchenko}, I.~A. and {De Donato}, C. and {De Santis}, C. and {De Simone}, N. and {Di Felice}, V. and {Formato}, V. and {Galper}, A.~M. and {Karelin}, A.~V. and {Koldashov}, S.~V. and {Koldobskiy}, S. and {Krutkov}, S.~Y. and {Kvashnin}, A.~N. and {Leonov}, A. and {Malakhov}, V. and {Marcelli}, L. and {Martucci}, M. and {Mayorov}, A.~G. and {Menn}, W. and {Merg{\'e}}, M. and {Mikhailov}, V.~V. and {Mocchiutti}, E. and {Monaco}, A. and {Mori}, N. and {Munini}, R. and {Osteria}, G. and {Palma}, F. and {Panico}, B. and {Papini}, P. and {Pearce}, M. and {Picozza}, P. and {Pizzolotto}, C. and {Ricci}, M. and {Ricciarini}, S.~B. and {Rossetto}, L. and {Sarkar}, R. and {Scotti}, V. and {Simon}, M. and {Sparvoli}, R. and {Spillantini}, P. and {Stozhkov}, Y.~I. and {Vacchi}, A. and {Vannuccini}, E. and {Vasilyev}, G.~I. and {Voronov}, S.~A. and {Yurkin}, Y.~T. and {Zampa}, G. and {Zampa}, N. and {Zverev}, V.~G.},
        title = "{Measurement of Boron and Carbon Fluxes in Cosmic Rays with the PAMELA Experiment}",
      journal = {The Astrophysical Journal},
         year = 2014,
        month = aug,
       volume = {791},
       number = {2},
          eid = {93},
        pages = {93},
          doi = {10.1088/0004-637X/791/2/93},
archivePrefix = {arXiv},
       eprint = {1407.1657},
 primaryClass = {astro-ph.HE},
       adsurl = {https://ui.adsabs.harvard.edu/abs/2014ApJ...791...93A}
}

@ARTICLE{Aguilar15,
       author = {{Aguilar}, M. and {Aisa}, D. and {Alpat}, B. and {Alvino}, A. and {Ambrosi}, G. and {Andeen}, K. and {Arruda}, L. and {Attig}, N. and {Azzarello}, P. and {Bachlechner}, A. and {Barao}, F. and {Barrau}, A. and {Barrin}, L. and {Bartoloni}, A. and {Basara}, L. and {Battarbee}, M. and {Battiston}, R. and {Bazo}, J. and {Becker}, U. and {Behlmann}, M. and {Beischer}, B. and {Berdugo}, J. and {Bertucci}, B. and {Bindi}, V. and {Bizzaglia}, S. and {Bizzarri}, M. and {Boella}, G. and {de Boer}, W. and {Bollweg}, K. and {Bonnivard}, V. and {Borgia}, B. and {Borsini}, S. and {Boschini}, M.~J. and {Bourquin}, M. and {Burger}, J. and {Cadoux}, F. and {Cai}, X.~D. and {Capell}, M. and {Caroff}, S. and {Casaus}, J. and {Castellini}, G. and {Cernuda}, I. and {Cerreta}, D. and {Cervelli}, F. and {Chae}, M.~J. and {Chang}, Y.~H. and {Chen}, A.~I. and {Chen}, G.~M. and {Chen}, H. and {Chen}, H.~S. and {Cheng}, L. and {Chou}, H.~Y. and {Choumilov}, E. and {Choutko}, V. and {Chung}, C.~H. and {Clark}, C. and {Clavero}, R. and {Coignet}, G. and {Consolandi}, C. and {Contin}, A. and {Corti}, C. and {Gil}, E. Cortina and {Coste}, B. and {Creus}, W. and {Crispoltoni}, M. and {Cui}, Z. and {Dai}, Y.~M. and {Delgado}, C. and {Della Torre}, S. and {Demirk{\"o}z}, M.~B. and {Derome}, L. and {Di Falco}, S. and {Di Masso}, L. and {Dimiccoli}, F. and {D{\'\i}az}, C. and {von Doetinchem}, P. and {Donnini}, F. and {Duranti}, M. and {D'Urso}, D. and {Egorov}, A. and {Eline}, A. and {Eppling}, F.~J. and {Eronen}, T. and {Fan}, Y.~Y. and {Farnesini}, L. and {Feng}, J. and {Fiandrini}, E. and {Fiasson}, A. and {Finch}, E. and {Fisher}, P. and {Formato}, V. and {Galaktionov}, Y. and {Gallucci}, G. and {Garc{\'\i}a}, B. and {Garc{\'\i}a-L{\'o}pez}, R. and {Gargiulo}, C. and {Gast}, H. and {Gebauer}, I. and {Gervasi}, M. and {Ghelfi}, A. and {Giovacchini}, F. and {Goglov}, P. and {Gong}, J. and {Goy}, C. and {Grabski}, V. and {Grandi}, D. and {Graziani}, M. and {Guandalini}, C. and {Guerri}, I. and {Guo}, K.~H. and {Haas}, D. and {Habiby}, M. and {Haino}, S. and {Han}, K.~C. and {He}, Z.~H. and {Heil}, M. and {Hoffman}, J. and {Hsieh}, T.~H. and {Huang}, Z.~C. and {Huh}, C. and {Incagli}, M. and {Ionica}, M. and {Jang}, W.~Y. and {Jinchi}, H. and {Kanishev}, K. and {Kim}, G.~N. and {Kim}, K.~S. and {Kirn}, Th. and {Korkmaz}, M.~A. and {Kossakowski}, R. and {Kounina}, O. and {Kounine}, A. and {Koutsenko}, V. and {Krafczyk}, M.~S. and {La Vacca}, G. and {Laudi}, E. and {Laurenti}, G. and {Lazzizzera}, I. and {Lebedev}, A. and {Lee}, H.~T. and {Lee}, S.~C. and {Leluc}, C. and {Li}, H.~L. and {Li}, J.~Q. and {Li}, J.~Q. and {Li}, Q. and {Li}, Q. and {Li}, T.~X. and {Li}, W. and {Li}, Y. and {Li}, Z.~H. and {Li}, Z.~Y. and {Lim}, S. and {Lin}, C.~H. and {Lipari}, P. and {Lippert}, T. and {Liu}, D. and {Liu}, H. and {Liu}, Hu and {Lolli}, M. and {Lomtadze}, T. and {Lu}, M.~J. and {Lu}, S.~Q. and {Lu}, Y.~S. and {Luebelsmeyer}, K. and {Luo}, F. and {Luo}, J.~Z. and {Lv}, S.~S. and {Majka}, R. and {Ma{\~n}{\'a}}, C. and {Mar{\'\i}n}, J. and {Martin}, T. and {Mart{\'\i}nez}, G. and {Masi}, N. and {Maurin}, D. and {Menchaca-Rocha}, A. and {Meng}, Q. and {Mo}, D.~C. and {Morescalchi}, L. and {Mott}, P. and {M{\"u}ller}, M. and {Nelson}, T. and {Ni}, J.~Q. and {Nikonov}, N. and {Nozzoli}, F. and {Nunes}, P. and {Obermeier}, A. and {Oliva}, A. and {Orcinha}, M. and {Palmonari}, F. and {Palomares}, C. and {Paniccia}, M. and {Papi}, A. and {Pauluzzi}, M. and {Pedreschi}, E. and {Pensotti}, S. and {Pereira}, R. and {Picot-Clemente}, N. and {Pilo}, F. and {Piluso}, A. and {Pizzolotto}, C. and {Plyaskin}, V. and {Pohl}, M. and {Poireau}, V. and {Putze}, A. and {Quadrani}, L. and {Qi}, X.~M. and {Qin}, X. and {Qu}, Z.~Y. and {R{\"a}ih{\"a}}, T. and {Rancoita}, P.~G. and {Rapin}, D. and {Ricol}, J.~S. and {Rodr{\'\i}guez}, I. and {Rosier-Lees}, S. and {Rozhkov}, A. and {Rozza}, D. and {Sagdeev}, R. and {Sandweiss}, J. and {Saouter}, P. and {Schael}, S. and {Schmidt}, S.~M. and {von Dratzig}, A. Schulz and {Schwering}, G. and {Scolieri}, G. and {Seo}, E.~S. and {Shan}, B.~S. and {Shan}, Y.~H. and {Shi}, J.~Y. and {Shi}, X.~Y. and {Shi}, Y.~M. and {Siedenburg}, T. and {Son}, D. and {Song}, J.~W. and {Spada}, F. and {Spinella}, F. and {Sun}, W. and {Sun}, W.~H. and {Tacconi}, M. and {Tang}, C.~P. and {Tang}, X.~W. and {Tang}, Z.~C. and {Tao}, L. and {Tescaro}, D. and {Ting}, Samuel C.~C. and {Ting}, S.~M. and {Tomassetti}, N. and {Torsti}, J. and {T{\"u}rko{\v{g}}lu}, C. and {Urban}, T. and {Vagelli}, V. and {Valente}, E. and {Vannini}, C. and {Valtonen}, E. and {Vaurynovich}, S. and {Vecchi}, M. and {Velasco}, M. and {Vialle}, J.~P. and {Vitale}, V. and {Vitillo}, S. and {Wang}, L.~Q. and {Wang}, N.~H. and {Wang}, Q.~L. and {Wang}, R.~S. and {Wang}, X. and {Wang}, Z.~X. and {Weng}, Z.~L. and {Whitman}, K. and {Wienkenh{\"o}ver}, J. and {Willenbrock}, M. and {Wu}, H. and {Wu}, X. and {Xia}, X. and {Xie}, M. and {Xie}, S. and {Xiong}, R.~Q. and {Xu}, N.~S. and {Xu}, W. and {Yan}, Q. and {Yang}, J. and {Yang}, M. and {Yang}, Y. and {Ye}, Q.~H. and {Yi}, H. and {Yu}, Y.~J. and {Yu}, Z.~Q. and {Zeissler}, S. and {Zhang}, C. and {Zhang}, J.~H. and {Zhang}, M.~T. and {Zhang}, S.~D. and {Zhang}, S.~W. and {Zhang}, X.~B. and {Zhang}, Z. and {Zheng}, Z.~M. and {Zhuang}, H.~L. and {Zhukov}, V. and {Zichichi}, A. and {Zimmermann}, N. and {Zuccon}, P. and {AMS Collaboration}},
        title = "{Precision Measurement of the Helium Flux in Primary Cosmic Rays of Rigidities 1.9 GV to 3 TV with the Alpha Magnetic Spectrometer on the International Space Station}",
      journal = {Physical Review Letters},
         year = 2015,
        month = Nov,
       volume = {115},
       number = {21},
          eid = {211101},
        pages = {211101},
          doi = {10.1103/PhysRevLett.115.211101},
       adsurl = {https://ui.adsabs.harvard.edu/abs/2015PhRvL.115u1101A}
}

@ARTICLE{Bell12,
       author = {{Schure}, K.~M. and {Bell}, A.~R. and {O'C Drury}, L. and {Bykov}, A.~M.},
        title = "{Diffusive Shock Acceleration and Magnetic Field Amplification}",
      journal = {Space Science Review },
         year = 2012,
        month = nov,
       volume = {173},
       number = {1-4},
        pages = {491-519},
          doi = {10.1007/s11214-012-9871-7},
archivePrefix = {arXiv},
       eprint = {1203.1637},
 primaryClass = {astro-ph.HE},
       adsurl = {https://ui.adsabs.harvard.edu/abs/2012SSRv..173..491S}
}

@article{Aartsen04,
  title = {Measurement of the cosmic ray energy spectrum with IceTop-73},
  author = {Aartsen, M. G. and Abbasi, R. and Abdou, Y. and Ackermann, M. and Adams, J. and Aguilar, J. A. and Ahlers, M. and Altmann, D. and Auffenberg, J. and Bai, X. and Baker, M. and Barwick, S. W. and Baum, V. and Bay, R. and Beatty, J. J. and Bechet, S. and Becker Tjus, J. and Becker, K.-H. and Benabderrahmane, M. L. and BenZvi, S. and Berghaus, P. and Berley, D. and Bernardini, E. and Bernhard, A. and Bertrand, D. and Besson, D. Z. and Binder, G. and Bindig, D. and Bissok, M. and Blaufuss, E. and Blumenthal, J. and Boersma, D. J. and Bohaichuk, S. and Bohm, C. and Bose, D. and B\"oser, S. and Botner, O. and Brayeur, L. and Bretz, H.-P. and Brown, A. M. and Bruijn, R. and Brunner, J. and Carson, M. and Casey, J. and Casier, M. and Chirkin, D. and Christov, A. and Christy, B. and Clark, K. and Clevermann, F. and Coenders, S. and Cohen, S. and Cowen, D. F. and Cruz Silva, A. H. and Danninger, M. and Daughhetee, J. and Davis, J. C. and De Clercq, C. and De Ridder, S. and Desiati, P. and de Vries, K. D. and de With, M. and DeYoung, T. and D\'{\i}az-V\'elez, J. C. and Dunkman, M. and Eagan, R. and Eberhardt, B. and Eisch, J. and Ellsworth, R. W. and Euler, S. and Evenson, P. A. and Fadiran, O. and Fazely, A. R. and Fedynitch, A. and Feintzeig, J. and Feusels, T. and Filimonov, K. and Finley, C. and Fischer-Wasels, T. and Flis, S. and Franckowiak, A. and Frantzen, K. and Fuchs, T. and Gaisser, T. K. and Gallagher, J. and Gerhardt, L. and Gladstone, L. and Gl\"usenkamp, T. and Goldschmidt, A. and Golup, G. and Gonzalez, J. G. and Goodman, J. A. and G\'ora, D. and Grandmont, D. T. and Grant, D. and Gro\ss{}, A. and Ha, C. and Haj Ismail, A. and Hallen, P. and Hallgren, A. and Halzen, F. and Hanson, K. and Heereman, D. and Heinen, D. and Helbing, K. and Hellauer, R. and Hickford, S. and Hill, G. C. and Hoffman, K. D. and Hoffmann, R. and Homeier, A. and Hoshina, K. and Huelsnitz, W. and Hulth, P. O. and Hultqvist, K. and Hussain, S. and Ishihara, A. and Jacobi, E. and Jacobsen, J. and Jagielski, K. and Japaridze, G. S. and Jero, K. and Jlelati, O. and Kaminsky, B. and Kappes, A. and Karg, T. and Karle, A. and Kelley, J. L. and Kiryluk, J. and Kl\"as, J. and Klein, S. R. and K\"ohne, J.-H. and Kohnen, G. and Kolanoski, H. and K\"opke, L. and Kopper, C. and Kopper, S. and Koskinen, D. J. and Kowalski, M. and Krasberg, M. and Krings, K. and Kroll, G. and Kunnen, J. and Kurahashi, N. and Kuwabara, T. and Labare, M. and Landsman, H. and Larson, M. J. and Lesiak-Bzdak, M. and Leuermann, M. and Leute, J. and L\"unemann, J. and Mac\'{\i}ias, O. and Madsen, J. and Maggi, G. and Maruyama, R. and Mase, K. and Matis, H. S. and McNally, F. and Meagher, K. and Merck, M. and Meures, T. and Miarecki, S. and Middell, E. and Milke, N. and Miller, J. and Mohrmann, L. and Montaruli, T. and Morse, R. and Nahnhauer, R. and Naumann, U. and Niederhausen, H. and Nowicki, S. C. and Nygren, D. R. and Obertacke, A. and Odrowski, S. and Olivas, A. and Omairat, A. and O'Murchadha, A. and Paul, L. and Pepper, J. A. and P\'erez de los Heros, C. and Pfendner, C. and Pieloth, D. and Pinat, E. and Posselt, J. and Price, P. B. and Przybylski, G. T. and R\"adel, L. and Rameez, M. and Rawlins, K. and Redl, P. and Reimann, R. and Resconi, E. and Rhode, W. and Ribordy, M. and Richman, M. and Riedel, B. and Rodrigues, J. P. and Rott, C. and Ruhe, T. and Ruzybayev, B. and Ryckbosch, D. and Saba, S. M. and Salameh, T. and Sander, H.-G. and Santander, M. and Sarkar, S. and Schatto, K. and Scheriau, F. and Schmidt, T. and Schmitz, M. and Schoenen, S. and Sch\"oneberg, S. and Sch\"onwald, A. and Schukraft, A. and Schulte, L. and Schulz, O. and Seckel, D. and Sestayo, Y. and Seunarine, S. and Shanidze, R. and Sheremata, C. and Smith, M. W. E. and Soldin, D. and Spiczak, G. M. and Spiering, C. and Stamatikos, M. and Stanev, T. and Stasik, A. and Stezelberger, T. and Stokstad, R. G. and St\"o\ss{}l, A. and Strahler, E. A. and Str\"om, R. and Sullivan, G. W. and Taavola, H. and Taboada, I. and Tamburro, A. and Tepe, A. and Ter-Antonyan, S. and Te\ifmmode \check{s}\else \v{s}\fi{}i\ifmmode \acute{c}\else \'{c}\fi{}, G. and Tilav, S. and Toale, P. A. and Toscano, S. and Unger, E. and Usner, M. and Vallecorsa, S. and van Eijndhoven, N. and Van Overloop, A. and van Santen, J. and Vehring, M. and Voge, M. and Vraeghe, M. and Walck, C. and Waldenmaier, T. and Wallraff, M. and Weaver, Ch. and Wellons, M. and Wendt, C. and Westerhoff, S. and Whitehorn, N. and Wiebe, K. and Wiebusch, C. H. and Williams, D. R. and Wissing, H. and Wolf, M. and Wood, T. R. and Woschnagg, K. and Xu, C. and Xu, D. L. and Xu, X. W. and Yanez, J. P. and Yodh, G. and Yoshida, S. and Zarzhitsky, P. and Ziemann, J. and Zierke, S. and Zoll, M.},
  collaboration = {IceCube Collaboration},
  journal = {Phys. Rev. D},
  volume = {88},
  issue = {4},
  pages = {042004},
  numpages = {15},
  year = {2013},
  month = {Aug},
  publisher = {American Physical Society},
  doi = {10.1103/PhysRevD.88.042004},
  url = {https://link.aps.org/doi/10.1103/PhysRevD.88.042004}
}

@article{apeli4,
title = {The KASCADE-Grande energy spectrum of cosmic rays and the role of hadronic interaction models},
journal = {Advances in Space Research},
volume = {53},
number = {10},
pages = {1456-1469},
year = {2014},
note = {Cosmic Ray Origins: Viktor Hess Centennial Anniversary},
ISSN = {0273-1177},
doi = {https://doi.org/10.1016/j.asr.2013.05.008},
url = {https://www.sciencedirect.com/science/article/pii/S0273117713002639},
author = {W.D. Apel and J.C. Arteaga-Velázquez and K. Bekk and M. Bertaina and J. Blümer and H. Bozdog and I.M. Brancus and E. Cantoni and A. Chiavassa and F. Cossavella and K. Daumiller and V. {de Souza} and F. {Di Pierro} and P. Doll and R. Engel and J. Engler and M. Finger and B. Fuchs and D. Fuhrmann and H.J. Gils and R. Glasstetter and C. Grupen and A. Haungs and D. Heck and J.R. Hörandel and D. Huber and T. Huege and K.-H. Kampert and D. Kang and H.O. Klages and K. Link and P. Łuczak and M. Ludwig and H.J. Mathes and H.J. Mayer and M. Melissas and J. Milke and B. Mitrica and C. Morello and J. Oehlschläger and S. Ostapchenko and N. Palmieri and M. Petcu and T. Pierog and H. Rebel and M. Roth and H. Schieler and S. Schoo and F.G. Schröder and O. Sima and G. Toma and G.C. Trinchero and H. Ulrich and A. Weindl and J. Wochele and M. Wommer and J. Zabierowski}
}

@article{apjac7049bib1,
  author={A. Abeysekara and R. Alfaro and C. Alvarez and others},
  title={},
  journal={ApJ},
  volume={871},
  year={2019},
  pages={96},
  url={https://doi.org/10.3847/1538-4357/aaf5cc},
  DOI={10.3847/1538-4357/aaf5cc}
}

@article{apjac7049bib2,
  author={A. Achterberg and R. D. Blandford and S. P. Reynolds},
  title={},
  journal={A \& A},
  volume={281},
  year={1994},
  pages={220}
}

@article{apjac7049bib64,
  author={Y. S. Yoon and H. S. Ahn and P. S. Allison and others},
  title={},
  journal={ApJ},
  volume={728},
  year={2011},
  pages={122},
  url={https://doi.org/10.1088/0004-637X/728/2/122},
  DOI={10.1088/0004-637X/728/2/122}
}

@article{apjac7049bib65,
  author={Y. S. Yoon and T. Anderson and A. Barrau and others},
  title={},
  journal={ApJ},
  volume={839},
  year={2017},
  pages={5},
  url={https://doi.org/10.3847/1538-4357/aa68e4},
  DOI={10.3847/1538-4357/aa68e4}
}

@article{apjac7049bib66,
  author={Q. Yuan and B.-Q. Qiao and Y.-Q. Guo and Y.-Z. Fan and X.-J. Bi},
  title={},
  journal={FrPhy},
  volume={16},
  year={2020},
  pages={24501},
  url={https://doi.org/10.1007/s11467-020-0990-4},
  DOI={10.1007/s11467-020-0990-4}
}

@Article{Biermann19,
AUTHOR = {Biermann, Peter L. and Kronberg, Philipp P. and Allen, Michael L. and Meli, Athina and Seo, Eun-Suk},
TITLE = {The Origin of the Most Energetic Galactic Cosmic Rays: Supernova Explosions into Massive Star Plasma Winds},
JOURNAL = {Galaxies},
VOLUME = {7},
YEAR = {2019},
NUMBER = {2},
ARTICLE-NUMBER = {48},
URL = {https://www.mdpi.com/2075-4434/7/2/48},
ISSN = {2075-4434},
DOI = {10.3390/galaxies7020048}
}

@article{Kirk99,
doi = {10.1088/0954-3899/25/8/201},
url = {https://dx.doi.org/10.1088/0954-3899/25/8/201},
year = {1999},
month = {aug},
publisher = {},
volume = {25},
number = {8},
pages = {R163},
author = {J G Kirk and  P Duffy},
title = {Particle acceleration and relativistic shocks},
journal = {Journal of Physics G: Nuclear and Particle Physics}
}

@article{ansh1,
title = {Study of general relativistic shocks and their propagation in neutron stars},
journal = {Journal of High Energy Astrophysics},
volume = {36},
pages = {36-47},
year = {2022},
issn = {2214-4048},
doi = {https://doi.org/10.1016/j.jheap.2022.07.005},
url = {https://www.sciencedirect.com/science/article/pii/S2214404822000404},
author = {Ritam Mallick and Anshuman Verma}
}

@article{ansh2,
    author = {Verma, Anshuman and Mallick, Ritam},
    title = "{Shock waves in (1+1-dimensional) curved space-time}",
    journal = {Monthly Notices of the Royal Astronomical Society},
    volume = {522},
    number = {4},
    pages = {4801-4814},
    year = {2023},
    month = {04},
    issn = {0035-8711},
    doi = {10.1093/mnras/stad1245},
    url = {https://doi.org/10.1093/mnras/stad1245}
}

@ARTICLE{Bonazzola_1993,
       author = {{Bonazzola}, S. and {Gourgoulhon}, E. and {Salgado}, M. and {Marck}, J.~A.},
        title = "{Axisymmetric rotating relativistic bodies: A new numerical approach for 'exact' solutions}",
      journal = {Astronomy \& Astrophysics (A\&A)},
         year = 1993,
        month = nov,
       volume = {278},
       number = {2},
        pages = {421-443},
       adsurl = {https://ui.adsabs.harvard.edu/abs/1993A&A...278..421B}
}

@ARTICLE{Baring_2012,
       author = {{Summerlin}, Errol J. and {Baring}, Matthew G.},
        title = "{Diffusive Acceleration of Particles at Oblique, Relativistic, Magnetohydrodynamic Shocks}",
      journal = {The Astrophysical Journal},
         year = 2012,
        month = jan,
       volume = {745},
       number = {1},
          eid = {63},
        pages = {63},
          doi = {10.1088/0004-637X/745/1/63},
archivePrefix = {arXiv},
       eprint = {1110.5968},
 primaryClass = {astro-ph.HE},
       adsurl = {https://ui.adsabs.harvard.edu/abs/2012ApJ...745...63S}
}

@ARTICLE{Liu-2024,
       author = {{Liu}, Siming},
        title = "{Evidence for extreme PeV cosmic ray acceleration from LHAASO}",
      journal = {Journal of High Energy Astrophysics},
         year = 2024,
        month = nov,
       volume = {44},
        pages = {116-122},
          doi = {10.1016/j.jheap.2024.09.006},
       adsurl = {https://ui.adsabs.harvard.edu/abs/2024JHEAp..44..116L}
}

@ARTICLE{arbutina_2021,
       author = {{Arbutina}, Bojan and {Zekovi{\'c}}, Vladimir},
        title = "{On the distribution function of suprathermal particles at collisionless shocks}",
      journal = {Journal of High Energy Astrophysics},
         year = 2021,
        month = nov,
       volume = {32},
        pages = {65-70},
          doi = {10.1016/j.jheap.2021.08.003},
archivePrefix = {arXiv},
       eprint = {2108.09085},
 primaryClass = {physics.plasm-ph},
       adsurl = {https://ui.adsabs.harvard.edu/abs/2021JHEAp..32...65A}
}

@article{DEXHEIMER_2017,
title = {What is the magnetic field distribution for the equation of state of magnetized neutron stars?},
journal = {Physics Letters B},
volume = {773},
pages = {487-491},
year = {2017},
issn = {0370-2693},
doi = {https://doi.org/10.1016/j.physletb.2017.09.008},
url = {https://www.sciencedirect.com/science/article/pii/S0370269317307062},
author = {V. Dexheimer and B. Franzon and R.O. Gomes and R.L.S. Farias and S.S. Avancini and S. Schramm}
}

@article{LEVAN201544,
title = {Swift discoveries of new populations of extremely long duration high energy transient},
journal = {Journal of High Energy Astrophysics},
volume = {7},
pages = {44-55},
year = {2015},
issn = {2214-4048},
doi = {https://doi.org/10.1016/j.jheap.2015.05.004},
url = {https://www.sciencedirect.com/science/article/pii/S2214404815000221},
author = {A.J. Levan},
}

@ARTICLE{singh_2019,
       author = {{Singh}, Shailendra and {Mallick}, Ritam},
        title = "{Time-like detonation in presence of magnetic field}",
      journal = {Laser and Particle Beams},
         year = 2019,
        month = mar,
       volume = {37},
       number = {1},
        pages = {30-37},
          doi = {10.1017/S0263034619000041},
       adsurl = {https://ui.adsabs.harvard.edu/abs/2019LPB....37...30S}
}

@ARTICLE{mallick_2014a,
       author = {{Mallick}, Ritam and {Schramm}, Stefan},
        title = "{Deformation of a magnetized neutron star}",
      journal = {Physical Review C (PRC) },
         year = 2014,
        month = apr,
       volume = {89},
       number = {4},
          eid = {045805},
        pages = {045805},
          doi = {10.1103/PhysRevC.89.045805},
archivePrefix = {arXiv},
       eprint = {1307.5185},
 primaryClass = {astro-ph.HE},
       adsurl = {https://ui.adsabs.harvard.edu/abs/2014PhRvC..89d5805M}
}

@ARTICLE{mallick_2014b,
       author = {{Mallick}, Ritam and {Schramm}, Stefan},
        title = "{Oblique magnetohydrodynamic shocks: Space-like and time-like characteristics}",
      journal = {Physical Review C (PRC) },
         year = 2014,
        month = feb,
       volume = {89},
       number = {2},
          eid = {025801},
        pages = {025801},
          doi = {10.1103/PhysRevC.89.025801},
       adsurl = {https://ui.adsabs.harvard.edu/abs/2014PhRvC..89b5801M}
}

@ARTICLE{Timmes_1999ApJS,
       author = {{Timmes}, F.~X. and {Arnett}, Dave},
        title = "{The Accuracy, Consistency, and Speed of Five Equations of State for Stellar Hydrodynamics}",
      journal = {The Astrophysical Journal},
         year = 1999,
        month = nov,
       volume = {125},
       number = {1},
        pages = {277-294},
          doi = {10.1086/313271},
       adsurl = {https://ui.adsabs.harvard.edu/abs/1999ApJS..125..277T}
}

@BOOK{anile,
       author = {{Anile}, Angelo Marcello},
        title = "{Relativistic fluids and magneto-fluids : with applications in astrophysics and plasma physics}",
         year = 1989,
       adsurl = {https://ui.adsabs.harvard.edu/abs/1989rfmw.book.....A}
}

@ARTICLE{de-hoffmann,
       author = {{de Hoffmann}, F. and {Teller}, E.},
        title = "{Magneto-Hydrodynamic Shocks}",
      journal = {Physical Review},
         year = 1950,
        month = nov,
       volume = {80},
       number = {4},
        pages = {692-703},
          doi = {10.1103/PhysRev.80.692},
       adsurl = {https://ui.adsabs.harvard.edu/abs/1950PhRv...80..692D}
}

@ARTICLE{taub,
       author = {{Taub}, A.~H.},
        title = "{Relativistic Rankine-Hugoniot Equations}",
      journal = {Physical Review},
         year = 1948,
        month = aug,
       volume = {74},
       number = {3},
        pages = {328-334},
          doi = {10.1103/PhysRev.74.328},
       adsurl = {https://ui.adsabs.harvard.edu/abs/1948PhRv...74..328T}
}

@ARTICLE{bugaev_2000,
       author = {{Bugaev}, K.~A. and {Gorenstein}, M.~I. and {Greiner}, W.},
        title = "{Discontinuities in Relativistic Hydrodynamics Across Space-like and Time-like Hypersurfaces}",
      journal = {arXiv e-prints},
         year = 2000,
        month = sep,
          eid = {nucl-th/0009035},
        pages = {nucl-th/0009035},
          doi = {10.48550/arXiv.nucl-th/0009035},
archivePrefix = {arXiv},
       eprint = {nucl-th/0009035},
 primaryClass = {nucl-th},
       adsurl = {https://ui.adsabs.harvard.edu/abs/2000nucl.th...9035B}
}

@ARTICLE{zhang_2014,
       author = {{Zhang}, Sun},
        title = "{Shock wave evolution and discontinuity propagation for relativistic superfluid hydrodynamics with spontaneous symmetry breaking}",
      journal = {Physics Letters B},
         year = 2014,
        month = feb,
       volume = {729},
        pages = {136-142},
          doi = {10.1016/j.physletb.2014.01.014},
       adsurl = {https://ui.adsabs.harvard.edu/abs/2014PhLB..729..136Z}
}

@ARTICLE{taub_1978,
       author = {{Taub}, A.~H.},
        title = "{Relativistic Fluid Mechanics}",
      journal = {Annual Review of Fluid Mechanics},
         year = 1978,
        month = jan,
       volume = {10},
        pages = {301-332},
          doi = {10.1146/annurev.fl.10.010178.001505},
       adsurl = {https://ui.adsabs.harvard.edu/abs/1978AnRFM..10..301T}
}

@ARTICLE{debaes_1997,
       author = {{Bandyopadhyay}, Debades and {Chakrabarty}, Somenath and {Pal}, Subrata},
        title = "{Quantizing Magnetic Field and Quark-Hadron Phase Transition in a Neutron Star}",
      journal = {Physical review letters},
         year = 1997,
        month = sep,
       volume = {79},
       number = {12},
        pages = {2176-2179},
          doi = {10.1103/PhysRevLett.79.2176},
archivePrefix = {arXiv},
       eprint = {astro-ph/9703066},
 primaryClass = {astro-ph},
       adsurl = {https://ui.adsabs.harvard.edu/abs/1997PhRvL..79.2176B}
}

@ARTICLE{prakash_2000,
       author = {{Broderick}, A. and {Prakash}, M. and {Lattimer}, J.~M.},
        title = "{The Equation of State of Neutron Star Matter in Strong Magnetic Fields}",
      journal = {The Astrophysical Journal},
         year = 2000,
        month = jul,
       volume = {537},
       number = {1},
        pages = {351-367},
          doi = {10.1086/309010},
archivePrefix = {arXiv},
       eprint = {astro-ph/0001537},
 primaryClass = {astro-ph},
       adsurl = {https://ui.adsabs.harvard.edu/abs/2000ApJ...537..351B}
}

@ARTICLE{ferraro_1954,
       author = {{Ferraro}, V.~C.~A.},
        title = "{On the Equilibrium of Magnetic Stars.}",
      journal = {The Astrophysical Journal},
         year = 1954,
        month = mar,
       volume = {119},
        pages = {407},
          doi = {10.1086/145838},
       adsurl = {https://ui.adsabs.harvard.edu/abs/1954ApJ...119..407F}
}

@ARTICLE{konno_1999,
       author = {{Konno}, K. and {Obata}, T. and {Kojima}, Y.},
        title = "{Deformation of relativistic magnetized stars}",
      journal = {Astronomy \& Astrophysics (A\&A)},
         year = 1999,
        month = dec,
       volume = {352},
        pages = {211-216},
          doi = {10.48550/arXiv.gr-qc/9910038},
archivePrefix = {arXiv},
       eprint = {gr-qc/9910038},
 primaryClass = {gr-qc},
       adsurl = {https://ui.adsabs.harvard.edu/abs/1999A&A...352..211K}
}

@article{rm_singh19,
  title = {Magnetized Taub adiabat and the $P T$ characteristics of magnetic neutron stars},
  author = {Mallick, Ritam and Singh, Shailendra},
  journal = {Phys. Rev. C},
  volume = {100},
  issue = {1},
  pages = {015801},
  numpages = {12},
  year = {2019},
  month = {Jul},
  publisher = {American Physical Society},
  doi = {10.1103/PhysRevC.100.015801},
  url = {https://link.aps.org/doi/10.1103/PhysRevC.100.015801}
}

@article{Mallick_2020,
    author = "Mallick, Ritam and Singh, Shailendra and Nandi, Rana",
    title = "{Maximum mass of hybrid star formed via shock induced phase transition in cold neutron stars}",
    eprint = "2012.04212",
    archivePrefix = "arXiv",
    primaryClass = "astro-ph.HE",
    doi = "10.1093/mnras/stab417",
    journal = "Mon. Not. Roy. Astron. Soc.",
    volume = "503",
    number = "4",
    pages = "4829--4837",
    year = "2021"
}

@article{Mishustin_2014,
    author = "Mishustin, Igor and Mallick, Ritam and Nandi, Rana and Satarov, Leonid",
    title = "{Phase transition in compact stars due to a violent shock}",
    eprint = "1410.8322",
    archivePrefix = "arXiv",
    primaryClass = "astro-ph.HE",
    doi = "10.1103/PhysRevC.91.055806",
    journal = "Phys. Rev. C",
    volume = "91",
    pages = "055806",
    year = "2015"
}

@ARTICLE{evangelia_2022,
       author = {{Ntormousi}, Evangelia and {Del Sordo}, Fabio and {Cantiello}, Matteo and {Ferrara}, Andrea},
        title = "{A closer look at supernovae as seeds for galactic magnetization}",
      journal = {Astronomy \& Astrophysics},
         year = 2022,
        month = Dec,
       volume = {668},
          eid = {L6},
        pages = {L6},
          doi = {10.1051/0004-6361/202245295},
archivePrefix = {arXiv},
       eprint = {2211.12355},
 primaryClass = {astro-ph.GA},
       adsurl = {https://ui.adsabs.harvard.edu/abs/2022A&A...668L...6N}
}

@ARTICLE{Donati-2002,
       author = {{Donati}, J. -F. and {Babel}, J. and {Harries}, T.~J. and {Howarth}, I.~D. and {Petit}, P. and {Semel}, M.},
        title = "{The magnetic field and wind confinement of {\ensuremath{\theta}}$^{1}$ Orionis C}",
      journal = {MNRAS},
         year = 2002,
        month = jun,
       volume = {333},
       number = {1},
        pages = {55-70},
          doi = {10.1046/j.1365-8711.2002.05379.x},
       adsurl = {https://ui.adsabs.harvard.edu/abs/2002MNRAS.333...55D}
}

@ARTICLE{Donati-2006,
       author = {{Donati}, J. -F. and {Howarth}, I.~D. and {Bouret}, J. -C. and {Petit}, P. and {Catala}, C. and {Landstreet}, J.},
        title = "{Discovery of a strong magnetic field on the O star HD 191612: new clues to the future of {\ensuremath{\theta}}$^{1}$ Orionis C}",
      journal = {MNRAS},
         year = 2006,
        month = jan,
       volume = {365},
       number = {1},
        pages = {L6-L10},
          doi = {10.1111/j.1745-3933.2005.00115.x},
archivePrefix = {arXiv},
       eprint = {astro-ph/0510395},
 primaryClass = {astro-ph},
       adsurl = {https://ui.adsabs.harvard.edu/abs/2006MNRAS.365L...6D}
}

@ARTICLE{Hubrig-2006,
       author = {{Hubrig}, S. and {Briquet}, M. and {Sch{\"o}ller}, M. and {De Cat}, P. and {Mathys}, G. and {Aerts}, C.},
        title = "{Discovery of magnetic fields in the {\ensuremath{\beta}}Cephei star {\ensuremath{\xi}}$^{1}$ CMa and in several slowly pulsating B stars$^{*}$}",
      journal = {MNRAS},
         year = 2006,
        month = jun,
       volume = {369},
       number = {1},
        pages = {L61-L65},
          doi = {10.1111/j.1745-3933.2006.00175.x},
archivePrefix = {arXiv},
       eprint = {astro-ph/0604283},
 primaryClass = {astro-ph},
       adsurl = {https://ui.adsabs.harvard.edu/abs/2006MNRAS.369L..61H}
}

@INPROCEEDINGS{Henrichs-2000,
       author = {{Henrichs}, H.~F. and {de Jong}, J.~A. and {Donati}, J. -F. and {Catala}, C. and {Wade}, G.~A. and {Shorlin}, S.~L.~S. and {Veen}, P.~M. and {Nichols}, J.~S. and {Kaper}, L.},
        title = "{The Magnetic Field of {\ensuremath{\beta}} Cep and the Be Phenomenon}",
    booktitle = {IAU Colloq. 175: The Be Phenomenon in Early-Type Stars},
         year = 2000,
       editor = {{Smith}, Myron A. and {Henrichs}, Huib F. and {Fabregat}, Juan},
       series = {Astronomical Society of the Pacific Conference Series},
       volume = {214},
        month = jan,
        pages = {324},
       adsurl = {https://ui.adsabs.harvard.edu/abs/2000ASPC..214..324H}
}

@inproceedings{yudin2009stars,
  title={Stars: Be, Polarization, Magnetic Fields},
  author={Yudin, R and Hubrig, S and Pogodin, M and others},
  booktitle={IAU Symposium},
  volume={259},
  pages={397},
  year={2009}
}

@incollection{pade_1994,
title = {Padé approximations},
series = {Handbook of Numerical Analysis},
publisher = {Elsevier},
volume = {3},
pages = {47-222},
year = {1994},
issn = {1570-8659},
doi = {https://doi.org/10.1016/S1570-8659(05)80016-X},
url = {https://www.sciencedirect.com/science/article/pii/S157086590580016X},
author = {Claude Brezinski and Jeannette {Van Iseghem}}
}

@article{Kuzur_2020,
    author = "Kuzur, Debojoti and Bhattacharyya, Rupamoy and Mallick, Ritam",
    title = "{Acceleration of charged particles in rotating magnetized star}",
    eprint = "2003.05197",
    archivePrefix = "arXiv",
    primaryClass = "astro-ph.HE",
    doi = "10.1088/1361-6471/aba9b0",
    journal = "J. Phys. G",
    volume = "47",
    number = "10",
    pages = "105203",
    year = "2020"
}

\appendix

\section{Conversion from HT to NI frame}\label{A1}
The shock planes of the HT and NI frames are coincident, as shown in Fig. \ref{f1:frame}. As done by Summelin and Baring \cite{Baring_2012} and Kirk and Heavens \cite{Kirk99}, to transform the Hoffmann-Teller (HT) frame to the Normal Incident Frame, we apply a Lorentz boost along the $y$-direction with boost velocity $v_1 = v_{ay}$. The corresponding Lorentz factor is given by:
\begin{equation}
    \Gamma_1 = \frac{1}{\sqrt{1 - v_1^2}}.
\end{equation}

The transformation equations for velocity components are:
\begin{align}
    v'_{ax} &= v_{ax} \Gamma_1, \\
    v'_{ay} &= 0, \\
    v'_{bx} &= \Gamma_1 \frac{v_{bx} - v_1^2 v_{bx}}{1 - \Gamma_1 v_1 v_{by}}, \\
    v'_{by} &= \frac{v'_{bx} v_{by} - v_{bx} v_1}{v_{bx}}.
\end{align}

Similarly, the transformation equations for the incident and reflected angles are:
\begin{align}
    \tan\theta'_a &= \frac{\tan\theta_a}{\Gamma_1}, \\
    \tan\theta'_b &= \frac{\tan\theta_b}{\Gamma_1}.
\end{align}

For the magnetic field transformation, the components transform as follows:
\begin{align}
    B'_{ax} &= B_{ax} \Gamma_1, \\
    B'_{ay} &= B_{ay}\\
    B'_{bx} &= B_{bx} \Gamma_1, \\
    B'_{by} &= B_{by}.
\end{align}
The $y$-component remains unchanged, while the $x$-component scales with $\Gamma_1$.These transformations ensure consistency in transitioning from the HT frame to the NI frame, preserving the physical properties necessary for shock analysis in relativistic magnetohydrodynamics. This method is useful because it removes all the imaginary and unphysical quantities in the superluminal shock. The shock smoothly transitions from the subluminal to the superluminal regime.

\end{document}